\documentclass[fleqn,usenatbib]{mnras}

\usepackage{newtxtext,newtxmath}

\usepackage[T1]{fontenc}

\DeclareRobustCommand{\VAN}[3]{#2}
\let\VANthebibliography\thebibliography
\def\thebibliography{\DeclareRobustCommand{\VAN}[3]{##3}\VANthebibliography}

\usepackage{graphicx}	
\usepackage{amsmath}	

\title{Confirming candidate self-lensing binary detections with radial velocity measurements}

\author[Adam McMaster et al.]{
Adam McMaster,$^{1}$\thanks{E-mail: a.mcmaster@soton.ac.uk}
Madeline-Mai Ward,$^{1}$
Hugh J. Dickinson,$^{2}$
Matthew J. Middleton,$^{1}$
D. J. Dixon,$^{3}$
\newauthor \;
Adam Ingram,$^{3,4}$
N. Castro Segura,$^{5}$
Simon Vaughan,$^{6}$
and Grzegorz Wiktorowicz$^{7,8}$
\\
$^{1}$Department of Physics and Astronomy, University of Southampton, Highfield, Southampton SO17 1BJ, UK\\
$^{2}$School of Physical Sciences, The Open University, Milton Keynes, MK7 6AA, UK\\
$^{3}$School of Mathematics, Statistics, and Physics, Newcastle University, Newcastle upon Tyne NE1 7RU, UK\\
$^{4}$Centre for Extragalactic Astronomy, Department of Physics, Durham University, South Road, Durham DH1 3LE, UK\\
$^{5}$Department of Physics, University of Warwick, Gibbet Hill Road, Coventry CV4 7AL, UK\\
$^{6}$School of Physics and Astronomy, University of Leicester, Leicester, LE1 7RH, UK\\
$^{7}$Astronomical Observatory of the University of Warsaw, Al. Ujazdowskie 4, 00-478 Warsaw, Poland\\
$^{8}$Nicolaus Copernicus Astronomical Center, Polish Academy of Sciences, Bartycka 18, 00-716, Warsaw, Poland
}

\date{Accepted XXX. Received YYY; in original form ZZZ}

\pubyear{\the\year{}}

\begin{document}
\label{firstpage}
\pagerange{\pageref{firstpage}--\pageref{lastpage}}
\maketitle

\begin{abstract}
    Compact objects in approximately edge-on binary orbits with stars are able to produce a self-lensing effect, amplifying light from their companion once per orbit. To date, five white dwarf self-lensing systems are known, while none are known containing black holes or neutron stars. Discovery of such systems is a potential method of finding quiescent black holes and neutron stars, most of which will be pristine and unchanged from their formation. Photometric detections can suffer from degeneracies in the lens mass and orbital parameters when the amplification and crossing time are not well constrained, while photometric searches produce too many false positives to allow radial velocity investigation of every candidate. Therefore we investigate the minimal radial velocity (RV) observational requirements to characterise such systems by simulating detectable self-lensing light curves, using those light curves to generate targeted RV observations, and fitting RV jointly with photometry. We evaluate the median relative error in the recovered lens mass and orbital period, alongside compact-object class recovery. We find that adding RV measurements substantially reduces parameter error, with compact-object class generally easier to determine than the exact lens mass. Parameter accuracy is driven mainly by number of RV observations, with RV precision playing a weaker role except for the shortest-period systems, with even a small number of RV measurements resulting in substantial improvement in accuracy. These results suggest that multi-stage RV follow-up, starting with a small number of observations of many candidates, can efficiently classify self-lensing candidates and identify high-priority systems for further investigation.
\end{abstract}




\section{Introduction}
There are expected to exist vast numbers of compact objects in binary systems in the Milky Way \citep[e.g.][]{shao_population_2019,wiktorowicz_noninteracting_2020,weller_mapping_2023}. The majority of these are not close enough to interact with their companions, making them challenging to detect. This is especially true of black holes, which would produce negligible electromagnetic emissions in such systems.

The detection of gravitational microlensing is a proven technique \citep{paczynski_gravitational_1986,paczynski_gravitational_1996,mao_astrophysical_2012}, in which the gravity of a foreground object (the \textit{lens}, which is often not visible) amplifies flux from a background object (the \textit{source}). This has been used to detect and study various classes of object, including exoplanets \citep{bond_ogle_2004,gaudi_microlensing_2012} and black holes \citep{sahu_isolated_2022,lam_isolated_2022}. Self-lensing is a special case of microlensing where the source and the lens are in a binary system together, in an orbit which is close to edge-on, producing a periodic microlensing amplification of the source once per orbit \citep{maeder_light_1973,agol_occultation_2002}. Self-lensing is only possible when the lens is a compact object. This means a detection of self-lensing (i.e. periodic microlensing) is \textit{de facto} a detection of a compact object.

A self-lensing search is an intriguing way to search for compact objects in binaries which would be undetectable by other means. To date, five self-lensing binaries have been confirmed, each containing a white dwarf lens, all of which have relatively close, short-period orbits \citep{kruse_koi-3278_2014,kawahara_discovery_2018,masuda_self-lensing_2019}. However, it is expected that stellar-compact object binaries should span the full range of compact object masses, as well as the full range of orbital configurations.

Photometry remains the best way to detect self-lensing binaries, and previous works have assessed the detectability of self-lensing binaries in various photometric surveys \citep{wiktorowicz_predicting_2021}. These results look promising, with modern all-sky surveys expected to imminently uncover a population of compact objects in self-lensing binaries, including the currently under-studied population of quiescent black holes.

\begin{figure*}
    \centering
    \includegraphics[width=\textwidth]{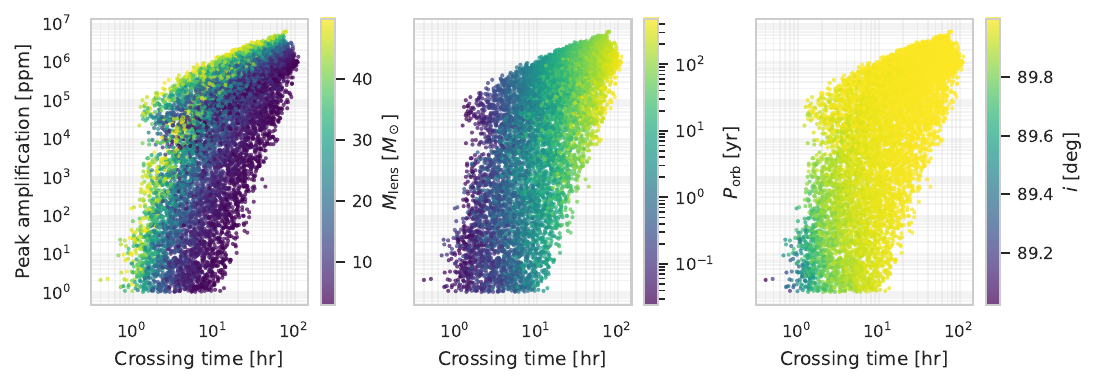}
    \caption{Relationship between self-lensing peak amplification, crossing time, lens mass, orbital period, and inclination for a Sun-like stellar companion.}
    \label{fig:self-lensing-observable-space}
\end{figure*}

While detection of self-lensing systems with photometry alone has been demonstrated, modelling of such systems and characterisation of the compact component can be subject to several degeneracies between the physical parameters which can make it challenging to accurately determine the mass of the lens in cases where the amplification due to self-lensing, the crossing time, or the stellar companion properties are not well known (see Figure~\ref{fig:self-lensing-observable-space} for an illustration of these relationships). For short-period binaries it should be possible to derive a photometric radial velocity (RV) by modelling Doppler boosting \citep{loeb_periodic_2003,zucker_beaming_2007,van_kerkwijk_observations_2010,faigler_photometric_2011}, but for systems with no detectable Doppler boosting effect, spectroscopic observations are the best way to solve for the binary's physical parameters \citep{kawahara_discovery_2018,yahalomi_mass_2019}. This is especially true for systems with the longest orbital periods, for which we may only be able to observe a single lensing event in a human lifetime (i.e. those with periods lasting decades, centuries, or longer).

In this work, we consider the question of how many spectroscopic RV measurements are required, and to what level of precision, in order to accurately retrieve the self-lensing system parameters as a function of those parameters. This includes systems with orbital periods on the order of centuries for which the orbital period can not be determined photometrically. This will inform observing strategies for follow-up of the large number of candidates expected to come from new surveys, allowing the most efficient use of resources. We also consider the usefulness of photometric-only fitting to guide RV observing strategy, developing a method of computing an RV observing schedule to maximise the separation between plausible solutions.

We will show that it is possible to accurately model very long period self-lensing binaries with a handful of RV measurements, even when periodicity can not be observed. In many cases, photometric-only fits can come close to determining the properties, removing the need for many RV measurements to retrieve the parameters accurately.

We begin by describing our methods for generating and fitting our simulated data in Section~\ref{sec:methods}, present and discuss the fitting results in Section~\ref{sec:results}, and conclude in Section~\ref{sec:conclusions}.

\section{Methods} \label{sec:methods}

\begin{figure*}
    \centering
    \includegraphics[width=\textwidth]{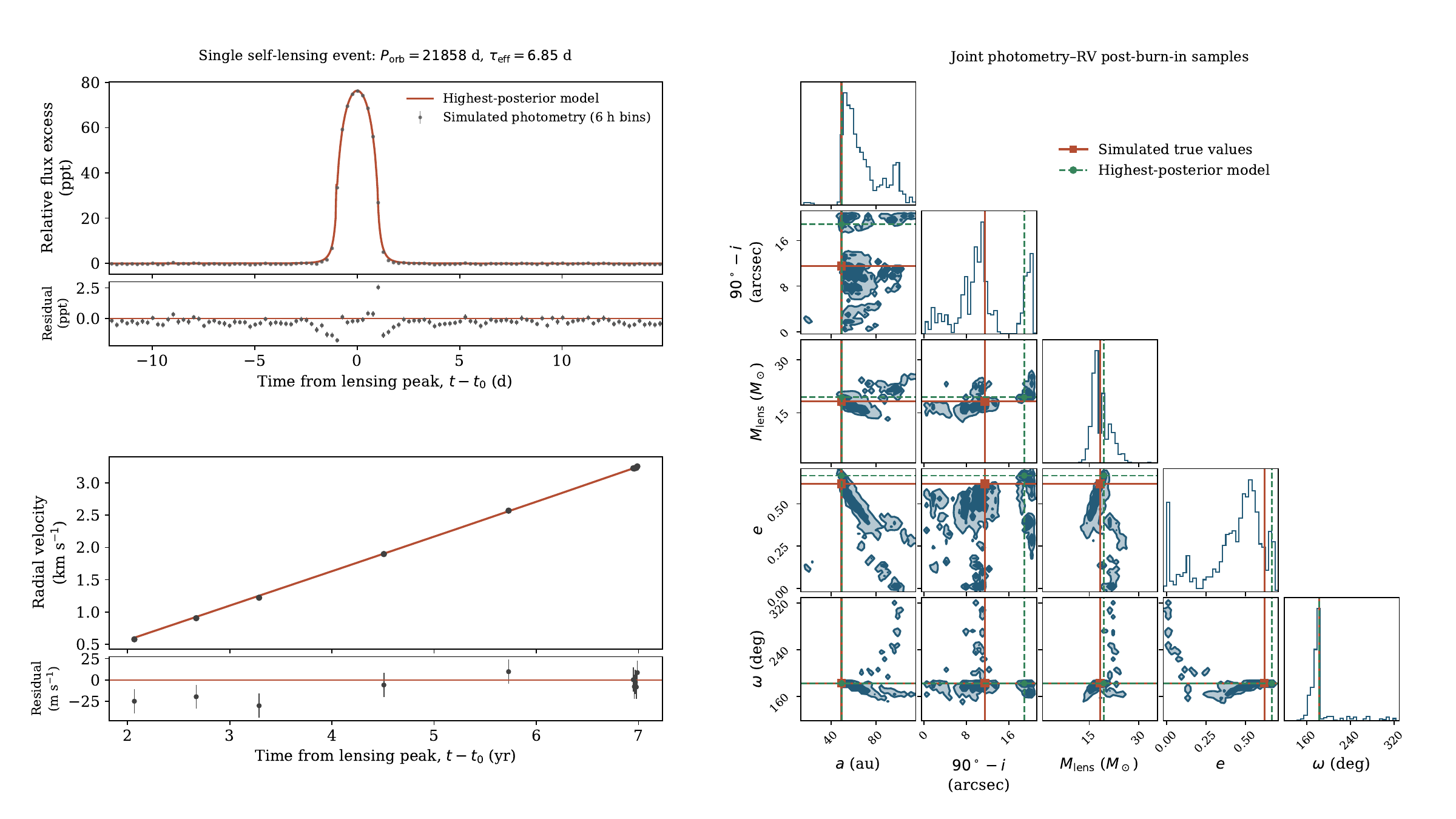}
    \caption{Example joint photometric--RV MCMC fit to a simulated self-lensing event. The simulated system has $M_\mathrm{lens}=18.29\,M_\odot$, $M_\star=15.21\,M_\odot$, $R_\star=8.83\,R_\odot$, $T_\mathrm{eff}=21032$ K, and distance $d=16.62$ kpc. Its orbital parameters are $a=49.32$ au, $e=0.620$, $i=89.99680^\circ$, and $\omega=182.79^\circ$, with $P_\mathrm{orb}=21858$ d and effective crossing time $\tau_\mathrm{eff}=6.85$ d. Left: the simulated photometry in 6-h bins and ten targeted RV measurements with instrumental accuracy $\sigma_\mathrm{ins}=10$ m s$^{-1}$, each with residuals beneath. The solid curves show the sample with the highest posterior value. Right: marginal distributions and joint 68 and 95 per cent contours from the post-burn-in joint-fit samples; red solid lines and square markers indicate the simulated true values, while green dashed lines and circular markers indicate the highest-posterior model. Inclination is shown as the angular departure from an edge-on orbit.}
    \label{fig:example-joint-fit}
\end{figure*}

We investigate the relationship between the number and precision of RV measurements and the relative error in recovered physical parameters as a function of lens mass, compact object type, and orbital period. We evaluate parameter accuracy using the median relative error, and separately consider recovery of broad compact-object class (i.e. distinguishing black holes from other compact objects even if the fitted mass remains imprecise).

When considering the recovery of compact object class, we adopt the following mass bins and object classes: $M_\mathrm{lens}<3\,M_\odot$ (white dwarfs or neutron stars); $3 \leq M_\mathrm{lens}/M_\odot < 5$ (lower mass gap); $5 \leq M_\mathrm{lens}/M_\odot < 20$ (lower mass black holes); $20 \leq M_\mathrm{lens}/M_\odot < 50$ (higher mass black holes). We separate the black hole mass ranges to assess how lens mass affects class recovery. A class is counted as recovered when the fitted lens mass falls in the same bin as the true lens mass.

Photometric detection of self-lensing signals is not the focus of the present work. Here we assume all of the simulated photometric lensing events have been detected and their precise timing has been determined. A robust method for carrying out such detections will be presented in a future work, including evaluation against self-lensing signals injected into real light curves. For now we focus only on the recovery of the physical properties of self-lensing systems which have already been detected, and how the RV observing strategy affects this.

As the basis of this work, we have generated a library of self-lensing photometric simulations covering a range of physical parameters, filtered to include only the simulated light curves which contain visible lensing, where the peak amplitude is larger than the photometric noise. We also exclude systems overflowing their Roche lobe and any systems where lensing is not geometrically possible. These simulations are described in Section~\ref{sec:phot-sims}.

An MCMC fit is then performed against the simulated photometry, to sample the posterior distribution of the self-lensing physical parameters and find distinct physical solutions for each light curve. We assume the stellar properties, mass, radius, effective temperature, and distance from Earth, are known. The set of distinct solutions is then used to compute an RV observing schedule using a variable number of observations (between three and ten observations) over a fixed five-year campaign, such that each schedule is optimised to distinguish between the photometric solutions.

Radial velocity measurements are then simulated for the computed observing schedule, using the same set of simulation parameters as the photometry (Section~\ref{sec:rv-sims}). Gaussian noise is added to simulate a range of spectrograph accuracies.

A second MCMC fit is then performed to sample the posterior distribution constrained jointly by the photometry and RV (Section~\ref{sec:fitting}). We then aggregate the median relative errors across observing parameters and self-lensing physical parameters to investigate how each observing schedule affects parameter retrieval.

Finally, we map the fitted outcomes onto previous population synthesis results to translate the estimated recovery error rates onto the expected real population of detectable self-lensing systems (Section~\ref{sec:pop-synth}).

\subsection{Self-lensing and photometric binary effects} \label{sec:phot-sims}

The photometric model used here is a modified version of COBALT-BLUE \citep{dixon_cobalt-blue_2026}, created while development of COBALT-BLUE was still in progress. Since the main focus of the present work is radial velocity scheduling and fitting, not the development of a full photometric forward model, we adopt simplified approximations to some aspects of the full model. Compared with COBALT-BLUE, our model adopts a local approximation to the eccentric self-lensing geometry and lower-order treatments of Doppler boosting and ellipsoidal variability, and we only consider linear limb darkening (\cite{dixon_cobalt-blue_2026} shows that linear limb darkening is sufficient for the present purposes). This work also adds a stellar noise model (Section~\ref{sec:gp}) and the spectroscopic radial-velocity component (Section~\ref{sec:rv-sims}) which are not included in COBALT-BLUE. A full mathematical description of the modified photometric model is given in Appendix~\ref{sec:phot-model}.

Simulations comprise a sample of self-lensing system parameters which is physically complete but not physically realistic (i.e. parameters are not sampled according to the frequency they are expected to occur in nature). This allows us to determine the ranges of parameters which produce self-lensing effects, independent of how common or uncommon particular parameter ranges are expected to be.

The photometric simulations comprise realistic simulated light curves with the same length and observing cadence as TESS light curves for each self-lensing system. When generating the simulations, the parameter values were sampled uniformly between the limits given in Table~\ref{tab:phot_sim_parameter_ranges}, with the exception of semi-major axis ($a$), eccentricity ($e$), and lens mass ($M_\mathrm{lens}$), which are sampled on a log scale to increase the number of samples at smaller values (the light curve morphology changes roughly in proportion to the relative change in these values; e.g. a 5 AU change in semi-major axis has a much larger effect going from 1 AU to 6 AU than for 100 AU to 105 AU).

\begin{table}
    \centering
    \caption{Parameter ranges used to generate the photometric simulation library.}
    \label{tab:phot_sim_parameter_ranges}
    \begin{tabular}{llll}
        \hline
        Parameter         & Description           & Range                & Spacing     \\
        \hline
        $a$               & Semi-major axis       & 0.1--200 AU          & Logarithmic \\
        $i$               & Orbital inclination   & 89.99--90.00$^\circ$ & Linear      \\
        $M_\mathrm{lens}$ & Lens mass             & 0.5--50 M$_\odot$    & Logarithmic \\
        $M_\star$         & Stellar mass          & 0.5--50 M$_\odot$    & Linear      \\
        $e$               & Eccentricity          & 0--0.999999          & Logarithmic \\
        $\omega$          & Periapsis phase angle & 0--360$^\circ$       & Linear      \\
        $d$               & Distance from Earth   & 0.1--100 kpc         & Linear      \\
        \hline
    \end{tabular}
\end{table}

Three photometric effects are included when generating the light curves: self-lensing, ellipsoidal variations, and Doppler boosting. At small values of $a$ the light curve is dominated by ellipsoidal variations and Doppler boosting (though lensing is often visible too); at large $a$ lensing is dominant, with ellipsoidal variation and Doppler boosting tending to zero as $a$ increases \citep{sorabella_modeling_2022}.

The peak amplitude and effective crossing time loosely correlate with $M_\mathrm{lens}$, $P_\mathrm{orb}$, and $i$ (Figure~\ref{fig:self-lensing-observable-space}). Since the amplitude and self-lensing event duration are directly measurable from the light curve, this can in principle allow the physical properties of the lens to be estimated. However, there is scatter in the relationship between these properties, since multiple lens configurations can produce similar amplifications and crossing times. Care must therefore be taken to account for this degeneracy when using these relationships (see discussion of MCMC chain initialisation in Section~\ref{sec:phot-fit}).

These effects all depend on the orbital phase at a given time during the simulation. The zero point for this is defined as the moment of maximum lensing magnification (i.e. the time of minimum projected separation between source and lens, $t_0$). The periapsis phase angle, $\omega$, is defined in terms of this zero point; i.e. the phase angle at which periastron passage occurs, relative to the position at $t_0$. Note that this is different from the convention used in COBALT-BLUE, which measures orbital phase relative to periastron passage.

All simulated light curves are 27 days long (the duration of a TESS sector). The peak time, $t_0$, is randomly selected to fall somewhere in the 27 day time window. An example light curve for a simulated self-lensing event can be seen in the top-left panel of Figure~\ref{fig:example-joint-fit}.

\subsection{Stellar noise} \label{sec:gp}


Gaussian processes (GPs) are designed to model smoothly varying correlated noise \citep{rasmussen_gaussian_2005}. In the exoplanet community, this has been highly successful in allowing exoplanet signals to be disentangled from stellar noise, though it does create significant computational expense which makes it impractical to scale GP-based detection to the millions of light curves being produced by current surveys.

In this work, we utilise a GP to generate realistic light curves with stellar noise, sampled according to the TESS observing cadence, instrumental sensitivity, and bandpass. We combine several kernels to model granulation and asteroseismic modes, based on the simulated stellar properties. We use \texttt{tinygp} \citep{foreman-mackey_dfmtinygp_2026} with quasi-separable simple harmonic oscillator (SHO) kernels.

Granulation is a form of aperiodic correlated noise due to convection cells at the stellar surface. Granulation appears as low-amplitude correlated noise; aperiodic in nature, it prefers certain timescales dictated by the size and age of the star \citep{kjeldsen_amplitudes_1995}. As derived
in \cite{foreman-mackey_exoplanet_2021}, we consider this a characteristic timescale of a damped stochastic process. We model such a process using a damped harmonic oscillator kernel with a fixed quality factor ($Q= \frac{1}{\sqrt{2}}$) parameter to create a damped random walk (DRW). This creates frequency-dependent broadband noise with a characteristic timescale but no dominant frequencies. The characteristic
timescale in the model $\tau$ then reflects the average lifetime of a single convective cell, while the amplitude scales with the number of granules contributing to the observable flux and the source-star surface gravity, $g_\star$ \citep{kjeldsen_amplitudes_1995,pereira_gaussian_2019,foreman-mackey_exoplanet_2021}. The granulation timescale,
$\tau_\mathrm{gran}$, for a source star of mass $M_\star$ and radius $R_\star$, goes as

\begin{equation}
    \tau_\mathrm{gran} \propto \frac{1}{g_\star}
    \propto \left(\frac{R_\star^2}{M_\star}\right).
\end{equation}

This model captures the temporal decorrelation of the granulation component light curve over time, appropriate for granulation, where surface features evolve on a finite timescale and exhibit only short-range temporal correlations \citep{aigrain_gaussian_2023}. This model has also been tested and applied in previous GP transit studies \citep{brewer_gaussian_2009,nicholson_quasi-periodic_2022}. Overall, the DRW kernel effectively models granulation noise without requiring high model complexity or prior knowledge of the star’s convective dynamics.

Asteroseismic oscillations are caused by the propagation of sound waves through the star's outer convective zone. \citep{kjeldsen_amplitudes_1995,chaplin_asteroseismology_2013}. These occur on the timescales of minutes to tens of hours, placing them directly into the observing window of light curves from instruments such as TESS. Again, we choose a damped harmonic oscillator kernel, but this time with a free quality factor to allow for the presence of a periodic component \citep{brewer_gaussian_2009,foreman-mackey_exoplanet_2021}. The frequency of the primary mode, given by the $\nu_\mathrm{seis}$ hyper-parameter, is proportional to the source-star surface gravity, $g_\star$, and the effective temperature, $T_\mathrm{eff}$, and therefore depends
on the stellar properties of the simulated system. We draw the oscillation quality factor independently for each simulated star from a broad log-uniform prior, $Q\in[100,10^4]$, chosen to represent coherent p modes.

\begin{equation}
    \nu_\mathrm{seis} \propto \frac{g_\star}{\sqrt{T_\mathrm{eff}}}
    \propto \frac{M_\star}{R_\star^2 \sqrt{T_\mathrm{eff}}},
\end{equation}
where $M_\star$, $R_\star$, and $T_\mathrm{eff}$ are the mass, radius, and effective
temperature of the source star.

In the GP implementation, these frequency scalings are evaluated from internally
consistent stellar relations rather than fitted independently for every light curve.
For a simulated stellar mass $M_\star$, we estimate the stellar radius and effective
temperature from main-sequence scaling relations. Let $\mathcal{R}$ be the dimensionless
mass-radius relation, $L(M_\star)$ the corresponding mass-luminosity relation, and
$\sigma_\mathrm{SB}$ the Stefan--Boltzmann constant; $M_\odot$ and $R_\odot$
are the solar mass and radius. Then
\begin{equation}
    R_\star = R_\odot \mathcal{R}(M_\star/M_\odot),
    \qquad
    T_\mathrm{eff} =
    \left(\frac{L(M_\star)}{4\pi \sigma_\mathrm{SB} R_\star^2}\right)^{1/4}.
\end{equation}
The asteroseismic frequency is then initialised using the usual
$\nu_\mathrm{max}$ scaling relation. Let $\nu_{\mathrm{max},\odot}$ be the
solar value of $\nu_\mathrm{max}$, and let $T_\odot$ be the solar effective
temperature. Then
\begin{equation}
    \nu_\mathrm{seis}
    =
    \nu_{\mathrm{max},\odot}
    \left(\frac{M_\star}{M_\odot}\right)
    \left(\frac{R_\star}{R_\odot}\right)^{-2}
    \left(\frac{T_\mathrm{eff}}{T_\odot}\right)^{-1/2},
\end{equation}
The granulation characteristic frequency is taken to be proportional to this value, adopting values around an empirical proportionality constant calibrated by \citep{kallinger_connection_2014},
\begin{equation}
    \nu_\mathrm{gran} = ~0.30\,\nu_\mathrm{seis}.
\end{equation}
The kernel hyperparameters are stored as $\log_{10}\nu$ in Hz and converted to angular frequency internally when constructing the SHO kernels.

To calculate the source brightness, we approximate the stellar spectrum as a blackbody with the radius and effective temperature derived above. The spectral flux density at Earth is
\begin{equation}
    F_\lambda = \pi B_\lambda(T_\mathrm{eff})\left(\frac{R_\star}{d}\right)^2,
\end{equation}
where $B_\lambda$ is the Planck function, $\lambda$ is wavelength, and $d$ is the source distance. We integrate the corresponding photon flux over the TESS bandpass, weighted by the instrumental response and detector quantum efficiency. Dividing this integral by the same integral for an AB zero-magnitude reference spectrum ($f_\nu=3631$ Jy) gives the ratio used to calculate the apparent TESS-band magnitude, $m_\mathrm{TESS}$, via $-2.5\log_{10}$ of that ratio. We then convert this magnitude to the expected detected source counts per cadence,
\begin{equation}
    N_\mathrm{TESS} = \frac{\Delta t}{1\,\mathrm{s}}\,10^{-0.4(m_\mathrm{TESS}-20.44)},
\end{equation}
using the instrumental magnitude zeropoint of 20.44 for a count rate of one electron per second and the simulated cadence $\Delta t=600$ s.

Kernel amplitudes are set in counts per cadence by scaling the fractional stellar-variability amplitudes with $N_\mathrm{TESS}$. The standard deviation of GP component $q$ is initialised as
\begin{equation}
    \sigma_q = f_q N_\mathrm{TESS},
\end{equation}
where $f_q$ is a fractional variability amplitude. We draw these amplitudes independently for each component and each simulated star from uniform distributions spanning 50--300 ppm for granulation and 100--1000 ppm for asteroseismic variability.

The same apparent magnitude is also used to set the TESS white-noise level. We linearly interpolate an empirical TESS noise model at $m_\mathrm{TESS}$ to obtain a fractional per-cadence uncertainty $\eta(m_\mathrm{TESS})$. For the GP covariance matrix $\Sigma_{ij}$ and the Kronecker delta $\delta_{ij}$, this gives a TESS white-noise standard deviation $\sigma_\mathrm{TESS}$ and diagonal covariance contribution
\begin{equation}
    \sigma_\mathrm{TESS} = \eta(m_\mathrm{TESS})N_\mathrm{TESS},
    \qquad
    \Sigma_{ij} \rightarrow \Sigma_{ij} + \sigma_\mathrm{TESS}^2 \delta_{ij}.
\end{equation}
This term is fixed by the simulated source brightness rather than fitted as an additional free GP jitter parameter.

The variability caused by magnetic effects has also been included in prior studies \citep{haywood_planets_2014,luger_starry_process_2021,foreman-mackey_exoplanet_2021}, as has the movement of star spots about the rotational axis of the star. However, such variations are often overpowered by more global variations, such as asteroseismic modes for distant stars, where surface structure cannot be resolved. We therefore assume that the jitter term is sufficient to account for magnetic contributions to the underlying stellar noise.

For a detailed review on modelling stellar variability through GPs, see \cite{aigrain_gaussian_2023}.

\subsection{Photometric MCMC fit} \label{sec:phot-fit}

Each simulated photometric light curve is fitted via MCMC using a random-walk Metropolis-Hastings sampler implemented with the BlackJAX package (128 chains, 1,000 total steps, including 200 burn-in steps). More modern MCMC samplers, such as Hamiltonian Monte Carlo, may be more effective or more efficient, but this requires a fully differentiable model. Our photometric model is differentiable in principle, however the current implementation around edges/parameter boundaries would result in discontinuous gradients in practice, potentially negatively affecting such samplers. This could be addressed in future work.

For these fits it is assumed we know the stellar physical parameters exactly, along with the relative lensing amplification (Appendix~\ref{sec:phot-model}) and effective crossing time, $\tau_\mathrm{eff}$ \citep{wiktorowicz_predicting_2021}. In reality there will be uncertainty in these values, but determining them is a separate problem which is out of scope for the present work. The photometric fit therefore represents an idealised best-case scenario for recovery of the physical properties from the photometry, eliminating the effects of the uncertainties in stellar models and allowing us to assess the RV observing schedule in isolation.

Light curves are binned according to their crossing time to minimise computation time while preserving the shape of the self-lensing peak (Table~\ref{tab:photometric-binning}). A limited number of bin widths was chosen for computational efficiency, as fits are run as vectorised batches on a GPU, requiring all the light curves in a batch to have the same length.

\begin{table}
    \centering
    \caption{Photometric bin widths used in the MCMC fits. The native simulated cadence is 10 minutes.}
    \label{tab:photometric-binning}
    \begin{tabular}{ll}
        \hline
        Effective crossing time, $\tau_\mathrm{eff}$ & Photometric bin width   \\
        \hline
        $\tau_\mathrm{eff} < 4$ h                    & 10 min (native cadence) \\
        $4$ h $\leq \tau_\mathrm{eff} \leq 1$ d      & 1 h                     \\
        $\tau_\mathrm{eff} > 1$ d                    & 6 h                     \\
        \hline
    \end{tabular}
\end{table}

We fit only the fundamental physical parameters of the system,
\begin{equation}
    \boldsymbol{\theta} = (a, i, M_\mathrm{lens}, e, \omega),
\end{equation}
with the stellar mass, stellar radius, effective temperature, distance, and lensing-peak time held fixed at their simulated values. Orbital period, $P_\mathrm{orb}$, impact parameter, maximum possible impact parameter, and effective lensing-event duration are derived from the fitted physical parameters rather than fitted directly. The orbital-period calculation is given in Appendix~\ref{sec:phot-model}.

Sampler chains are initialised using an estimated parameter distribution calculated from $\tau_\mathrm{eff}$ and $A_\mathrm{SL}$ (Appendix~\ref{sec:phot-model}). Similar to Figure~\ref{fig:self-lensing-observable-space}, parameters are calculated in the region around the observed values, and then these are clustered by the parameters in $\boldsymbol{\theta}$ using $k$-means clustering to find distinct groupings of estimated parameters. The central point of each cluster is used to initialise the sampler chains. This ensures the random walk thoroughly explores the area of the parameter space with the highest likelihood given the known amplification and crossing time, without constraining it from exploring beyond those regions.

We use a uniform prior within the corresponding parameter limits in Table~\ref{tab:phot_sim_parameter_ranges}, with negative infinity probability outside these limits. We also require trial systems to satisfy the self-lensing geometry condition $b \leq b_\mathrm{max}$, where $b$ is the impact parameter in units of the Einstein radius and $b_\mathrm{max}$ is its maximum value for self-lensing. Their expressions are given in Appendix~\ref{sec:phot-model}.

At the end of the MCMC run, we analyse the joint posterior distribution to identify distinct candidate solutions, accounting for degeneracies in which multiple self-lensing configurations can produce similar light curves. Examining the marginal posterior of each parameter separately is insufficient because the fitted parameters are correlated and only certain combinations represent physically plausible solutions. We therefore apply a heuristic clustering procedure that combines farthest-first traversal \citep{gonzalez_clustering_1985} with Lloyd’s $k$-means refinement algorithm \citep{lloyd_least_1982}. This is applied in a scaled representation of the joint parameter space, with the periodic phase represented using its sine and cosine. The number of clusters, $k$, is determined heuristically: candidate cluster centres are added using farthest-first traversal until no sample lies beyond a specified minimum separation from the existing centres, subject to a maximum of $k=8$. The centres are then refined using $k$-means, and clusters containing less than 3 per cent of the posterior samples are merged into the nearest retained cluster. The upper limit of eight clusters, the minimum centre separation, and the minimum cluster size of 3 per cent were chosen empirically from inspection of a random sample of fits, balancing the representation of distinct solutions against the computational cost of constructing the RV observing schedule. Because clustering is used only to construct the observing schedule and does not enter the subsequent parameter fitting or error calculation directly, we do not interpret the resulting clusters as formal estimates of the number of posterior modes.

\subsection{Radial velocity calculation} \label{sec:rv-sims}

The RV simulations use the same physical parameters and phase convention as the photometric simulations (Section~\ref{sec:phot-sims}). We define the orbital phase relative to $t_0$ as the true orbital longitude, using the true anomaly $\nu(t)$,
\begin{equation}
    \phi(t) = \omega + \nu(t),
\end{equation}
where $\nu(t)$ is measured relative to periapsis. The phase satisfies $\phi(t_0)=0$, and the true anomaly is calculated from Kepler's equation as described in Appendix~\ref{sec:phot-model}.

Let $\mu_g=G(M_\mathrm{lens}+M_\star)$ be the standard gravitational parameter, and $h=[\mu_g a(1-e^2)]^{1/2}$ be the specific angular momentum of the relative orbit. The source-star barycentric radial and tangential orbital-velocity components are
\begin{align}
    v_{\mathrm{rad},\star}(t) & =
    \frac{M_\mathrm{lens}}{M_\mathrm{lens}+M_\star}
    \frac{\mu_g}{h}
    e\sin\nu(t),
    \nonumber                     \\
    v_{\mathrm{tan},\star}(t) & =
    \frac{M_\mathrm{lens}}{M_\mathrm{lens}+M_\star}
    \frac{\mu_g}{h}
    \left[1 + e\cos\nu(t)\right].
\end{align}
The simulated line-of-sight radial velocity is then
\begin{equation}
    v_r(t) =
    \left[
        v_{\mathrm{tan},\star}(t)\sin\phi(t)
        -
        v_{\mathrm{rad},\star}(t)\cos\phi(t)
        \right]\sin i.
\end{equation}
This reduces to the circular-orbit convention $v_r(t)\propto\sin\phi(t)$ when $e=0$, but for eccentric orbits the radial component means the RV at $t_0$ is not generally zero.

We do not attempt to simulate realistic spectropic observations, which would require accounting for various instrumental factors including spectrograph stability, spectral resolution, and signal-to-noise ratio. Instead we add varying amounts of instrumental noise ($\sigma_\mathrm{ins} \in \{0.1, 1, 10, 100, 1000\}$ m/s) to simulate the total uncertainty in the radial velocity measurements that would result from the different instrumental effects. The values chosen for $\sigma_\mathrm{ins}$ span the orders of magnitude of expected RV accuracy of spectrographs in service today, from ESPRESSO \citep[][$\sigma_\mathrm{ins} = 0.1$ m/s]{pepe_espresso_2021} to APO ARCES \citep[][$\sigma_\mathrm{ins} = 0.5$ km/s]{gaulme_kic_2022}. We also add an additional fixed Gaussian noise to represent stochastic stellar variability \citep[][$\sigma_\star = 10$ m/s]{wright_radial_2005}. For simplicity, we do not include proper motion or heliocentric motion, as these would be corrected before fitting real observations.

An example of the radial velocity measurements for a simulated self-lensing system can be seen in the bottom-left panel of Figure~\ref{fig:example-joint-fit}.

\subsection{Targeted radial velocity scheduling} \label{sec:scheduling}

The simulated RV curves start at randomised times between 1 year and 10 years from the end of the photometry. This is to simulate a realistic delay in identifying candidates in the photometry and organising follow-up of those candidates. For this work we assume each observing campaign will last five years, and we trial varying numbers of RV observations, $3 \leq N_\mathrm{obs} \leq 10$. We do not attempt to simulate fully realistic observing (with observing seasons, seeing, varying weather conditions, etc.) as we are mainly interested in measuring the effects of instrument selection and number of observations.

The distinct solutions derived in Section~\ref{sec:phot-fit} are used to compute a schedule for the radial velocity observations, in which observations are deliberately timed to distinguish between the solutions and maximise the information gained about the binary orbit.

First, in the event that the photometric fit only found one plausible solution, the orbital period of that solution is checked to determine whether a lensing event will occur during the simulated observing window. If so, three observations are scheduled around this event, one at $t_0$, and two at $t_0 \pm \delta_t$. $\delta_t$ is chosen by drawing random samples from the photometric MCMC posterior and choosing the smallest value of $\delta_t$ where the change in RV is larger than $5 \sigma_\mathrm{ins}$ for $>90\%$ of samples, assuming $\omega = \pi$; i.e. the smallest time offset where a change in RV can be reliably measured given the specified value of $\sigma_\mathrm{ins}$. This triplet of observations provides information about $e$ and $\omega$: if $v_r(t_0) \neq 0$, then $e > 0$; if $v_r(t_0) = 0$, then either $e = 0$ or $\omega = 0$ (or both) and the value of $v_r(t_0 \pm \delta_t)$ provides information about these values.

If there are multiple plausible photometric MCMC solutions, the predicted RV values for each of them over the observing campaign are calculated, and the difference in RV values for each pair of solutions is calculated (each ``solution" is a cluster of posterior samples, so the separation is actually the squared difference between the mean RVs of each cluster, normalised by the error). The final set of observation times is chosen by iteratively selecting the times which maximise the minimum separation between solutions, for a minimum of three observations and a maximum of $\lceil N_\mathrm{obs}/2 \rceil$.

Any remaining observations are scheduled to minimise the gaps between observations. For a circular orbit, this is equivalent to limiting the gaps in orbital phase between observations, maximising the orbital phase coverage (within the constraint of the length of the observing campaign), and minimising the variances in fitted model parameters later derived from the observations \citep{koen_how_2024}.

The bottom-left panel of Figure~\ref{fig:example-joint-fit} shows the scheduled RV measurements for a simulated self-lensing system. There are ten RV observations in total, unevenly spaced so as to effectively distinguish between solutions.

\subsection{Joint photometric-RV MCMC fit} \label{sec:fitting}

Each simulated pair of photometric light curves and RV curves is then fitted via MCMC using the same setup as in Section~\ref{sec:phot-fit}, using the final chain state from the photometric fit as the initial state for the joint fit. Photometry is binned as per the first fit, while RV data are not binned. The Metropolis-Hastings random-walk proposal scale is reduced based on the number of RV measurements, when RV data are present, or otherwise on the number of photometric points, to allow the sampler to efficiently explore the narrower posterior where more data are present.

For a trial parameter vector $\boldsymbol{\theta}$, let $v_\mathrm{mod}(t_k;\boldsymbol{\theta})$ be the model RV at the $k$th RV observation time, and let $N_\mathrm{RV}$ be the number of RV measurements. The RV contribution to the fit statistic is
\begin{equation}
    Q_\mathrm{RV}(\boldsymbol{\theta}) =
    -\sum_{k=1}^{N_\mathrm{RV}}
    \left[
        \frac{v_k - v_\mathrm{mod}(t_k;\boldsymbol{\theta})}
        {\sigma_{\mathrm{RV},k}}
        \right]^2,
\end{equation}
where $v_k$ and $\sigma_{\mathrm{RV},k}$ are the simulated RV measurement and uncertainty. For the light curve, let $f_j$ and $\sigma_{\mathrm{LC},j}$ be the binned photometric flux and uncertainty at the $j$th binned time, let $f_\mathrm{mod}(t_j;\boldsymbol{\theta})$ be the corresponding model flux, and let $N_\mathrm{LC}$ be the number of fitted light-curve bins. The light-curve contribution is
\begin{equation}
    Q_\mathrm{LC}(\boldsymbol{\theta}) =
    -\sum_{j=1}^{N_\mathrm{LC}}
    \left[
        \frac{f_j - f_\mathrm{mod}(t_j;\boldsymbol{\theta})}
        {\sigma_{\mathrm{LC},j}}
        \right]^2.
\end{equation}
The quantity sampled by the MCMC is then the sum of these terms plus the parameter-prior term, $\ln \pi(\boldsymbol{\theta})$,
\begin{equation}
    \ln p(\boldsymbol{\theta}\mid\mathrm{data}) =
    \ln \pi(\boldsymbol{\theta}) +
    Q_\mathrm{RV}(\boldsymbol{\theta}) +
    Q_\mathrm{LC}(\boldsymbol{\theta}).
\end{equation}
The reported fitted parameters are posterior summaries rather than single maximum-posterior samples: after burn-in, we retain the 256 samples with the highest finite posterior values for each fit and take the mean of the fitted parameters, using a circular mean for the periapsis phase angle.

An example joint photometric-RV fit is shown in Figure~\ref{fig:example-joint-fit}. The corner plot in the right hand panel shows that the true solution is well described by the posterior distribution with the highest posterior model being close to the true values for $a$, $M_\mathrm{lens}$, $e$, and $\omega$, with only $i$ fitting to an incorrect value.

\subsection{Predicting recovery of real self-lensing systems} \label{sec:pop-synth}

In order to assess how the RV observing strategy would affect the recovery of the physical parameters of a physically realistic sample of self-lensing systems, we seek to combine our estimates of the recoverability of the physical properties of self-lensing systems with the detectability estimates of \cite{wiktorowicz_predicting_2021}. While the present work deliberately assesses a sample which is \textit{not} physically realistic, \cite{wiktorowicz_predicting_2021} uses population synthesis to generate a physically plausible distribution of physical properties, as the basis for simulating the total number of detectable systems observed by each photometric survey.

In the present work, we do not simulate a physically representative distribution of parameters, because our primary aim is to explore the ability to recover those parameters regardless of how common a given configuration will be. In other words, we want to simulate very rare configurations as thoroughly as more common ones. We are also concerned only with characterisation of lenses/recovery of physical parameters via spectroscopic RV measurements after detection in photometry, while \cite{wiktorowicz_predicting_2021} investigates photometric detectability.

Still, it is useful to derive from our simulations an estimate of how well we can recover the properties of a more realistic sample of self-lensing systems. To this end we build a nearest-neighbour emulator ($k=25$) to map the population synthesis results onto the fitting/recovery results for our radial velocity simulations.

We select population-synthesis entries marked as detectable and draw a sample of 10,000 entries with replacement, with selection probabilities proportional to their population-synthesis weights. This incorporates the relative frequencies represented by those weights into the sampled population, so each sampled entry contributes equally to the final statistics.

The emulator estimates the recovery probability for each population-synthesis system from the fitted outcomes of nearby systems in our simulation library. Here $i$ indexes our simulations and $j$ indexes the population-synthesis systems. We assign one scaled feature vector to each individual system in either sample, using its true physical parameters:
\begin{equation}
    \mathbf{x} =
    \left(
    \frac{\log_{10} M_\mathrm{lens} - \mu_M}{s_M},
    \frac{\log_{10} P_\mathrm{orb} - \mu_P}{s_P},
    \frac{e - \mu_e}{s_e},
    \frac{b - \mu_b}{s_b}
    \right).
\end{equation}
For each of the four coordinates, the centring constant $\mu$ is the median and the scale factor $s$ is the difference between the 84th and 16th percentiles, calculated from the combined simulation library and detected population-synthesis sample. For mass and period, these statistics are calculated after taking the logarithm. The same constants are applied to every system in both samples. This transformation shifts the origin and rescales each axis, placing both samples in a common dimensionless coordinate system and preventing parameters with larger numerical spreads from dominating the neighbour search.

For a population-synthesis system $j$, we calculate the Euclidean distance to each simulation $i$,
\begin{equation}
    d_{ij} = \lVert\mathbf{x}^{\mathrm{sim}}_i - \mathbf{x}^{\mathrm{pop}}_j\rVert.
\end{equation}
Thus, a vector describes an individual system, while a distance describes a pair of systems; the common centring constants cancel when taking the difference. We select the $k=25$ simulations with the smallest distances. The indices of the closest simulations form the set $\mathcal{N}_k(j)$, which can differ for each population-synthesis system. We assign these neighbours inverse-distance weights,
\begin{equation}
    w_{ij} =
    \frac{(d_{ij} + \epsilon)^{-1}}
    {\sum_{\ell \in \mathcal{N}_k(j)} (d_{\ell j} + \epsilon)^{-1}},
    \quad i \in \mathcal{N}_k(j),
\end{equation}
where $\epsilon = 10^{-12}$ avoids division by zero, and the weights sum to unity over the selected neighbours. For each observing strategy (specified by observing baseline, number of RV measurements, and RV accuracy), let $q_i$ denote the compact-object class recovery outcome of simulation $i$. We set $q_i=1$ if the fitted compact-object class matches the true class and $q_i=0$ otherwise.

The emulated class recovery probability for population-synthesis system $j$ is the weighted mean of its neighbours' class recovery outcomes,
\begin{equation}
    \hat{p}_{j} = \sum_{i \in \mathcal{N}_k(j)} w_{ij} q_i.
\end{equation}
The final class recovery fraction is the arithmetic mean of these emulated probabilities over the sampled population,
\begin{equation}
    \bar{p} = \frac{1}{N_\mathrm{pop}} \sum_{j=1}^{N_\mathrm{pop}} \hat{p}_j,
\end{equation}
where $N_\mathrm{pop}$ is the number of sampled population-synthesis entries.

We also emulate the relative error in lens mass. For simulation $i$ and the specified observing strategy, we define the relative error in fitted lens mass as $r_i=|M_{\mathrm{lens},i}^{\mathrm{fit}}-M_{\mathrm{lens},i}^{\mathrm{true}}|/M_{\mathrm{lens},i}^{\mathrm{true}}$. We then interpolate these errors using the same neighbour weights,
\begin{equation}
    \hat{r}_j =
    \sum_{i \in \mathcal{N}_k(j)} w_{ij} r_i.
\end{equation}
The reported median relative error is the median of the $\hat{r}_j$ values for sampled entries within each true lens-mass bin.

We note that the survey-specific detection yields we use here come from Table~2 in \cite{wiktorowicz_predicting_2021}, which is an optimistic upper limit including events for which only one photometric detection is made (i.e. one photometric point in the light curve falls within the lensing event). In reality, detections would require multiple photometric points to estimate $t_0$ and $\tau_\mathrm{eff}$.

\section{Results and discussion} \label{sec:results}

\begin{figure*}
    \centering
    \includegraphics[width=\textwidth]{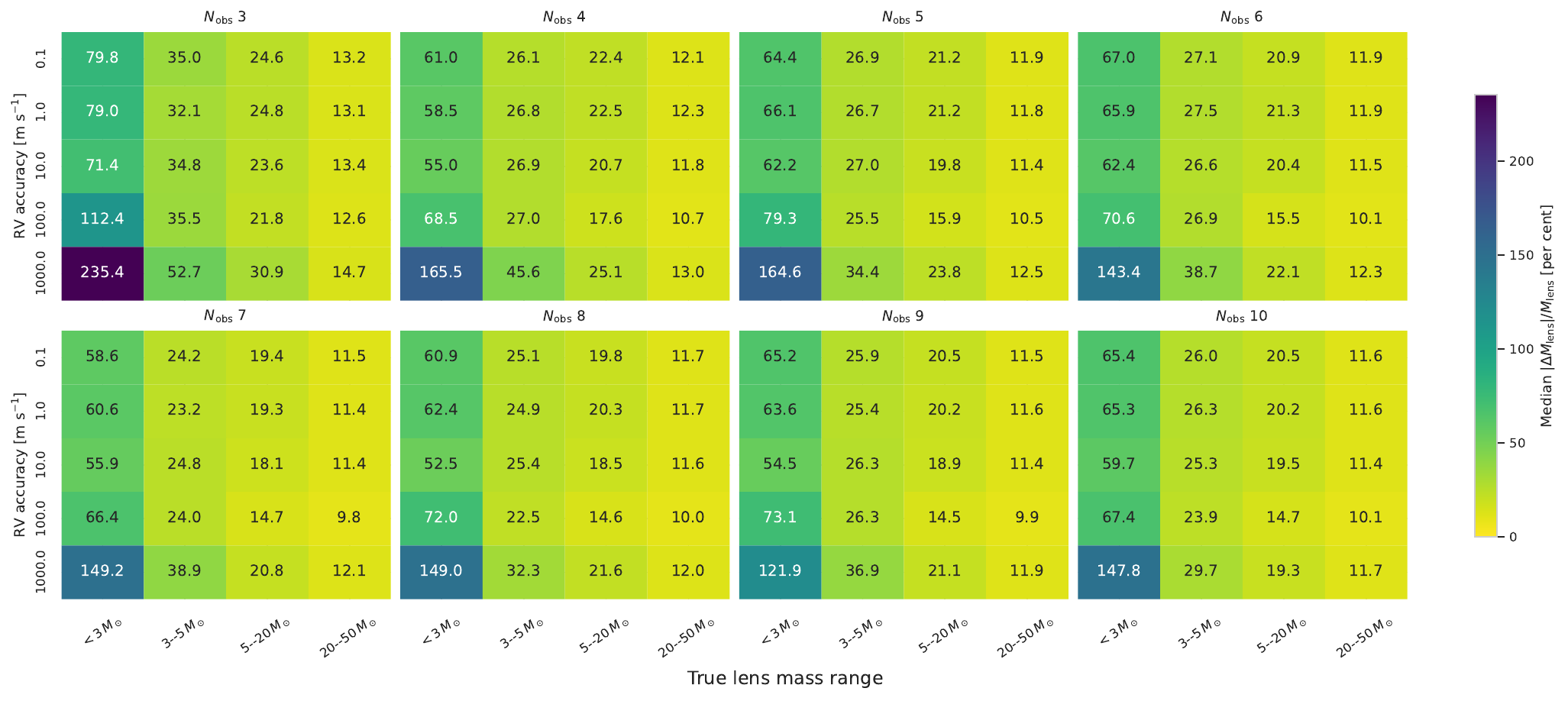}
    \caption{Median lens mass relative error as a function of the true lens mass. Cell values give the median $|\Delta M_\mathrm{lens}|/M_\mathrm{lens}$ in per cent for each fitted subset and observing strategy.}
    \label{fig:lens-mass-relative-error-by-lens-mass}
\end{figure*}

\begin{figure*}
    \centering
    \includegraphics[width=\textwidth]{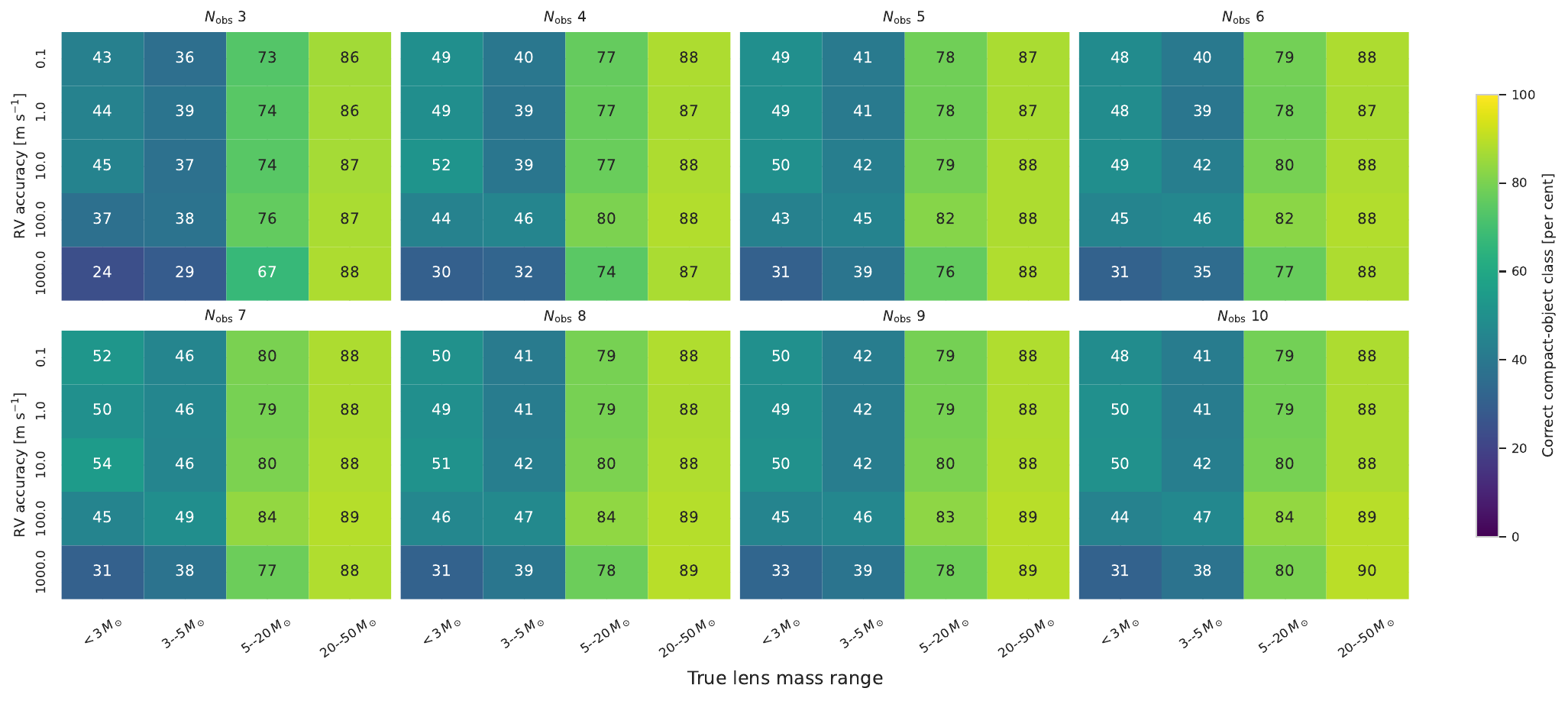}
    \caption{Compact-object class recovery fraction as a function of the true lens mass. A system is counted as recovered when the fitted mass falls in the same compact-object mass bin as the true mass, using $M_\mathrm{lens}<3\,M_\odot$ for white dwarf or neutron-star lenses, $3 \leq M_\mathrm{lens}/M_\odot < 5$ for the mass gap, $5 \leq M_\mathrm{lens}/M_\odot < 20$ for lower-mass black holes, and $20 \leq M_\mathrm{lens}/M_\odot < 50$ for higher-mass black holes.}
    \label{fig:lens-mass-class-recovery-by-lens-mass}
\end{figure*}

We evaluate the fits based on their ability to reproduce the true simulated parameters, with particular interest in lens mass and orbital period. The following subsections evaluate the fitting results in terms of the median relative error in lens mass and orbital period. We also evaluate recovery of the correct compact-object class regardless of the accuracy of the fitted mass, considering if we can distinguish black holes from white dwarfs and neutron stars, as this distinction affects planning for further follow-up observations. We then evaluate the efficacy of fitting photometry alone versus fitting RV alone or jointly fitting photometry and RV, assessing how the number of RV measurements affects the relative error. Finally we use these results to emulate recovery rates for the synthesised population from \cite{wiktorowicz_predicting_2021}, to estimate the relative errors in parameters for the real population of self-lensing systems.

It is worth noting that these results represent an ideal best case scenario for obtaining follow-up observations. We aim to assess a representative range of observing schedules, rather than rigorously simulating realistic observations. We do not account for observing seasons, we assume every target is always observable, we do not simulate weather or bad seeing, and we assume only a simple model of noise in the RV measurements (varying amounts of Gaussian noise to simulate spectrograph accuracy, ranging from sub-metre per second accuracy up to kilometre per second accuracy, to represent the range of spectrograph facilities currently available), plus a fixed 10 m/s Gaussian stellar noise. We also assume the exact time of the lensing peak ($t_0$) and the properties of the stellar companion (mass, radius, effective temperature, and distance from Earth) are known with no uncertainty; in practice these would all be subject to some amount of uncertainty.

\subsection{Lens mass relative error}

Figure~\ref{fig:lens-mass-relative-error-by-lens-mass} shows that after performing joint photometric/RV fitting, the results for higher lens masses generally have lower median relative errors. The number of observations has only a modest effect on the recovery of lens mass. For lower lens masses, the RV accuracy ($\sigma_\mathrm{ins}$) is an important factor, with $\sigma_\mathrm{ins} = 100 \text{ m s}^{-1}$ and $1 \text{ km s}^{-1}$ accuracies producing substantially worse results. Interestingly, better RV accuracy does not always result in better recovery of lens mass; for lens masses $> 5 \text{ M}_\odot$, $\sigma_\mathrm{ins} = 100 \text{ m s}^{-1}$ is better, while for the lowest mass lenses 10 m~s$^{-1}$ is best. The presumptive explanation for this is that a $\sigma_\mathrm{ins}$ smaller than the simulated stellar noise can confound the fitting procedure. Future work could incorporate a GP-based noise model into the RV likelihood function to improve recovery further.

\subsection{Recovering object class}
While the ultimate goal of our observing campaign will be to accurately measure the lens mass and other physical properties of the population of self-lensing systems, we anticipate having to grapple with many orders of magnitude more false positives than true self-lensing systems. This situation will therefore require an informed, systematic way of triaging and prioritising targets for follow-up. The volume of false positives is anticipated to be so large that we will want to obtain as few observations of each target as possible in order to characterise what type of compact object the lens is (or, indeed, whether each target is even a self-lensing system to begin with). In the present work we assume every target is \textit{some} kind of self-lensing system, but even with that assumption, some classes of object will be more astrophysically interesting than others, and some will require more or fewer observations in order to confirm. It is therefore informative to know whether we can accurately predict the \textit{class} of the lens, even in cases where we can not accurately predict the mass. Being able to do so will allow us to allocate resources accordingly when we plan follow-up observations.

Figure~\ref{fig:lens-mass-class-recovery-by-lens-mass} shows how well the object class is recovered as a function of the number of observations when performing a joint photometric/RV fit. A system with a fitted mass over 5 M$_\odot$ can often be identified as a black hole even when its fitted mass has a large error. This can inform our priorities when planning observations; e.g. we can choose to obtain additional observations of targets with fitted masses that fall in the 3--5 M$_\odot$ the mass gap, or we can treat higher mass systems as confirmed with fewer observations. We find that the number of observations has only a small effect on the recovery of the object class, while RV accuracy is more important for lower mass systems but not for high mass systems. This suggests a staged observing strategy would be appropriate, where lower accuracy observations are first obtained for all targets, with higher accuracy observations reserved for those which seem to be lower mass.

\subsection{Orbital period relative error}

The orbital period strongly influences the shape of the RV curve. In the circular-orbit limit, the RV is zero at the lensing event and reaches an extremum half an orbit later; for eccentric systems, eccentricity and periapsis phase shift the detailed RV shape and extrema. When we can fix the orbital phase at a specific time (as we can when lensing is observed), obtaining an RV curve, even over a partial orbit, can tightly constrain the orbital period, eccentricity, and periapsis phase angle even for very long orbits.

\begin{figure*}
    \centering
    \includegraphics[width=\textwidth]{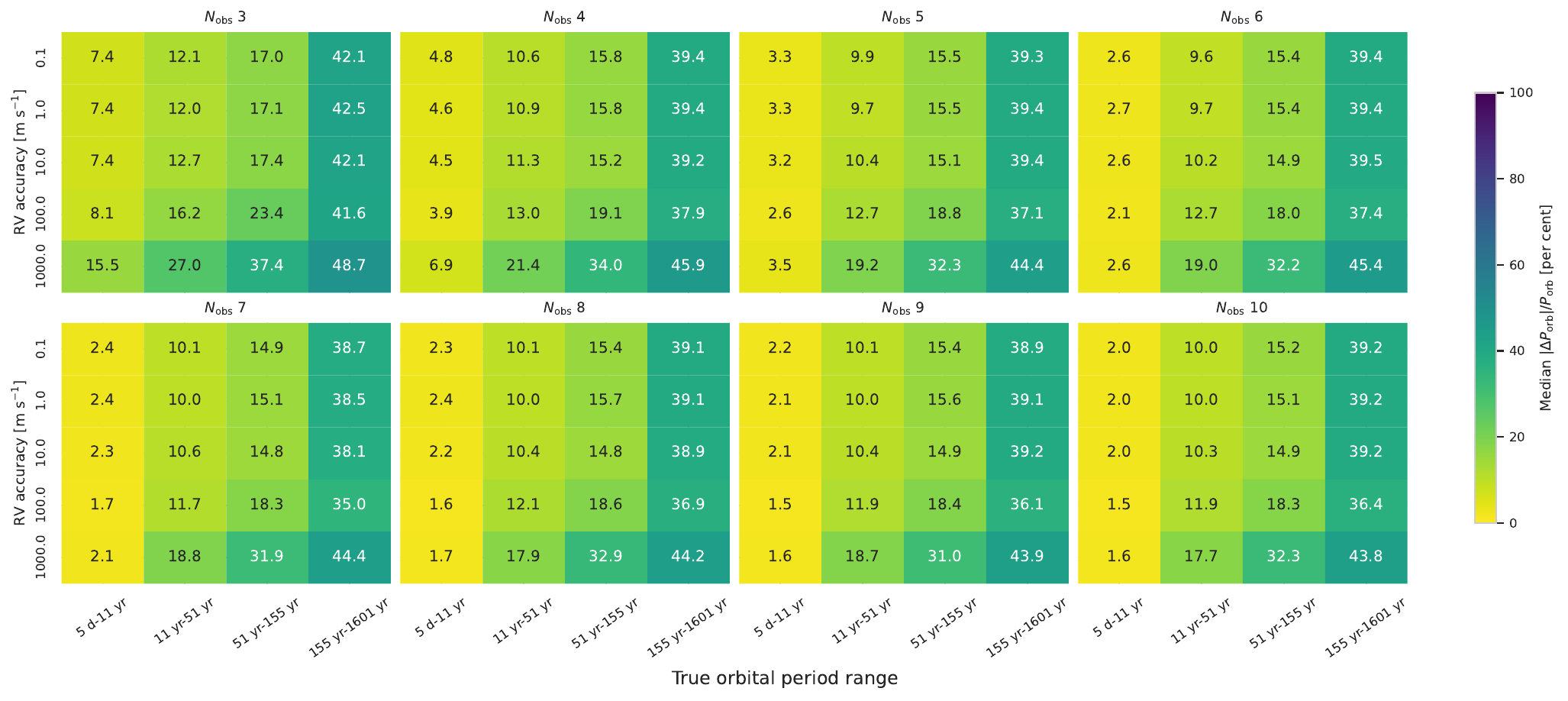}
    \caption{Median orbital-period relative error as a function of the true orbital period. Cell values give the median $|\Delta P_\mathrm{orb}|/P_\mathrm{orb}$ in per cent for each fitted subset and observing strategy.}
    \label{fig:orbital-period-relative-error-by-period}
\end{figure*}

Figure~\ref{fig:orbital-period-relative-error-by-period} shows the median relative error in orbital period as a function of number of observations, RV accuracy, and orbital period. Shorter orbital periods are recovered with smaller errors than longer ones; this is especially true for the shortest bin, where many systems will complete multiple orbits during the observing window.

\subsection{Photometry-only and RV-only vs joint fits and $N_\text{RV}$}

The simulations are generated on the assumption that we photometrically detect a candidate self-lensing event and then follow up with RV measurements some time later. The joint photometry and RV fits use the photometric peak time and the targeted schedules described above. For comparison, we also carried out RV-only fits using independently generated, evenly spaced observations and fit the RV data in isolation, with a free orbital phase and no photometric information.

\begin{figure*}
    \centering
    \includegraphics[width=\textwidth]{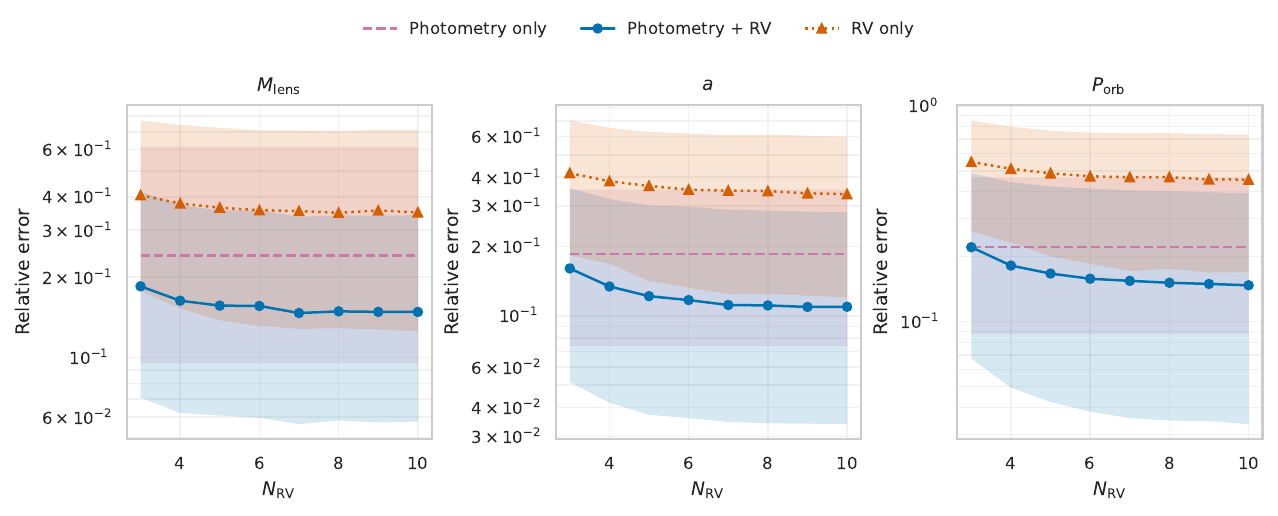}
    \caption{Comparison between the photometry-only pre-fit accuracy, RV-only fit accuracy, and joint photometry plus RV fit accuracy. The RV-only fits use an evenly spaced observing schedule and no photometric information; the photometry+RV fits use the photometric peak time and targeted RV schedule. Separate panels show the fitted parameters. Points show the median relative error in each parameter as a function of the number of RV measurements, and shaded regions show the interquartile range.}
    \label{fig:prefit-improvement-by-n-rv}
\end{figure*}

Figure~\ref{fig:prefit-improvement-by-n-rv} shows the median relative error in the recovered values of $M_\mathrm{lens}$, $a$, and $P_\mathrm{orb}$ as a function of the number of RV measurements. In all three cases, an RV-only fit performs worse than a photometry-only fit, and a joint fit has the lowest relative error. The median relative error in RV-only and joint fit converges after 5-6 observations.

\begin{figure*}
    \centering
    \includegraphics[height=0.4\textheight]{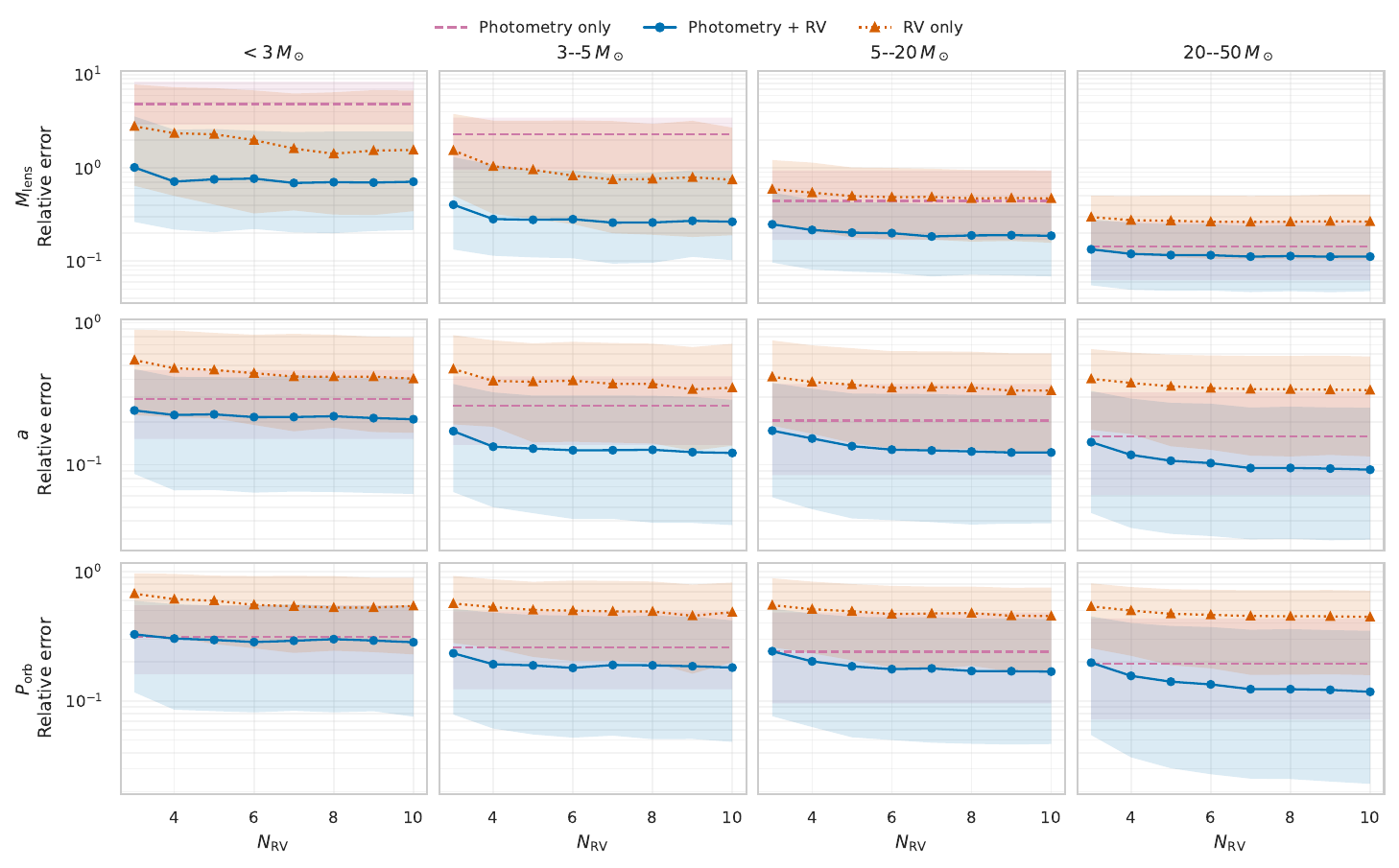}
    \caption{Comparison between photometry-only pre-fit accuracy, RV-only fit accuracy, and joint photometry plus RV fit accuracy, separated by true lens-mass range. RV-only fits use an evenly spaced observing schedule and no photometric information; photometry+RV fits use the photometric peak time and targeted RV schedule. Rows show fitted parameters and columns show true lens-mass ranges. Points show the median relative error as a function of the number of RV measurements, and shaded regions show the interquartile range.}
    \label{fig:prefit-improvement-by-lens-mass-range}
\end{figure*}

Figure~\ref{fig:prefit-improvement-by-lens-mass-range} shows the median improvement of fitted parameters broken down by the lens mass. The relative error is comparable for each $M_\mathrm{lens}$ bin when fitting $a$ and $P_\mathrm{orb}$ (with a modest improvement for larger masses). However, when fitting $M_\mathrm{lens}$ we find that higher lens masses can be fitted more accurately than lower masses. The improvement in the relative error when doing a joint fit is smaller for higher lenses, and in fact the RV-only fit is worse than the photometry-only fit at higher masses, suggesting RV measurements are the most useful for lower mass lenses. For the highest lens masses, fitting photometry alone results in recovered lens masses almost as accurate as a joint fit.

\begin{figure*}
    \centering
    \includegraphics[width=\textwidth]{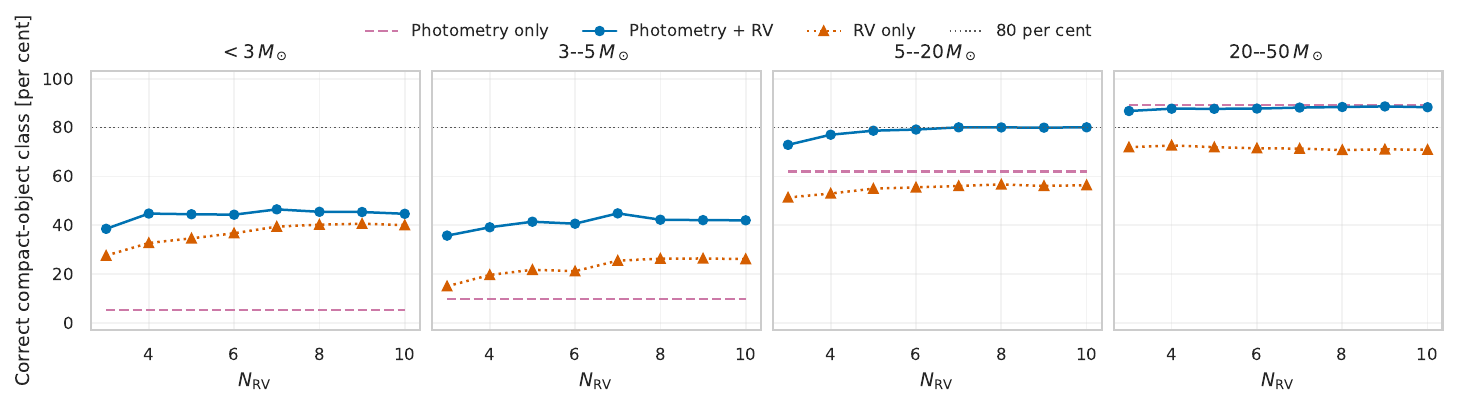}
    \caption{Compact-object class recovery for the photometry-only pre-fit, RV-only fit, and joint photometry plus RV fit, separated by true lens-mass range. RV-only fits use an evenly spaced observing schedule and no photometric information; photometry+RV fits use the photometric peak time and targeted RV schedule. A system is counted as correct when the fitted lens mass falls in the same broad compact-object class as the true mass, using $M_\mathrm{lens}<3\,M_\odot$ for white dwarf or neutron-star lenses, $3 \leq M_\mathrm{lens}/M_\odot < 5$ for the mass gap, $5 \leq M_\mathrm{lens}/M_\odot < 20$ for lower-mass black holes, and $20 \leq M_\mathrm{lens}/M_\odot < 50$ for higher-mass black holes. Points show the percentage of fitted systems in the correct class as a function of the number of RV measurements; the horizontal dotted line marks 80 per cent.}
    \label{fig:prefit-compact-object-class-by-lens-mass-range}
\end{figure*}

Figure~\ref{fig:prefit-compact-object-class-by-lens-mass-range} shows the percentage of systems which fall into the correct compact object mass bin regardless of the accuracy of the fitted mass. There is a clear trend where higher mass objects can be more accurately determined to be such, even if their exact mass can not be constrained. A higher mass black hole creates such a prominent lensing curve that it can be identified more easily, whereas neutron stars and white dwarfs are more ambiguous. It may be surprising that the photometry-only fits look so much worse for low mass lenses in Figure~\ref{fig:prefit-compact-object-class-by-lens-mass-range} than in Figure~\ref{fig:prefit-improvement-by-lens-mass-range}; this is because Figure~\ref{fig:prefit-improvement-by-lens-mass-range} shows the relative error, with a median error close to 500\% for the smallest mass bin. An error of that scale is more likely to result in lower mass systems falling in the wrong mass bin than larger mass systems (which have wider bins in this analysis). This is important astrophysically, as it makes it harder to distinguish neutron stars from mass gap black holes with photometry alone; for such candidates it is therefore more important to obtain RV measurements in order to characterise the nature of the lens.

\subsection{Predicted recovery rates}

We used the nearest-neighbour emulator to estimate the recovery rates for parameters based on the population synthesis-based parameter distribution. Recovery rates are the predicted percentages of detectable systems we could recover, calculated by emulating the results of our RV simulations across the physically-motivated population-synthesis distribution used in \cite{wiktorowicz_predicting_2021}. These recovery rates are independent of photometric survey in this model, but still depend on the RV observing strategy.

\begin{figure*}
    \centering
    \includegraphics[width=\textwidth]{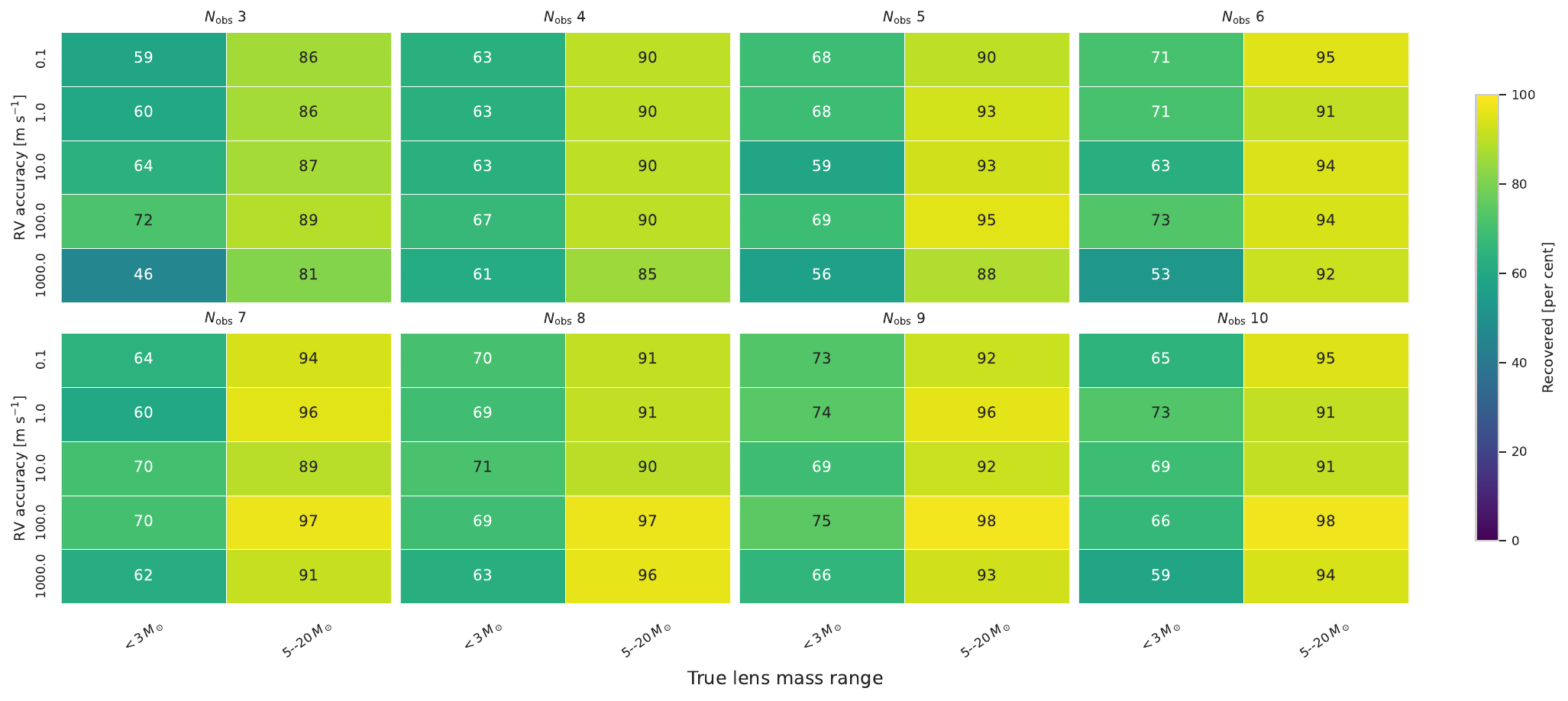}
    \caption{Compact-object class recovery percentage, scaled by population-synthesis results, for each observing strategy and instrumental RV accuracy. Columns within each panel give bins in true lens mass; the results are aggregated over orbital period. The population-synthesis sample contains no objects in the omitted lens-mass ranges.}
    \label{fig:population-weighted-rv-followup-heatmaps-counts-class}
\end{figure*}

Figure~\ref{fig:population-weighted-rv-followup-heatmaps-counts-class} shows the recovery rates for compact-object class. Class recovery spans roughly 60--100\% depending on number of RV observations, with higher mass lenses being recovered more successfully than lower mass.

\begin{figure*}
    \centering
    \includegraphics[width=\textwidth]{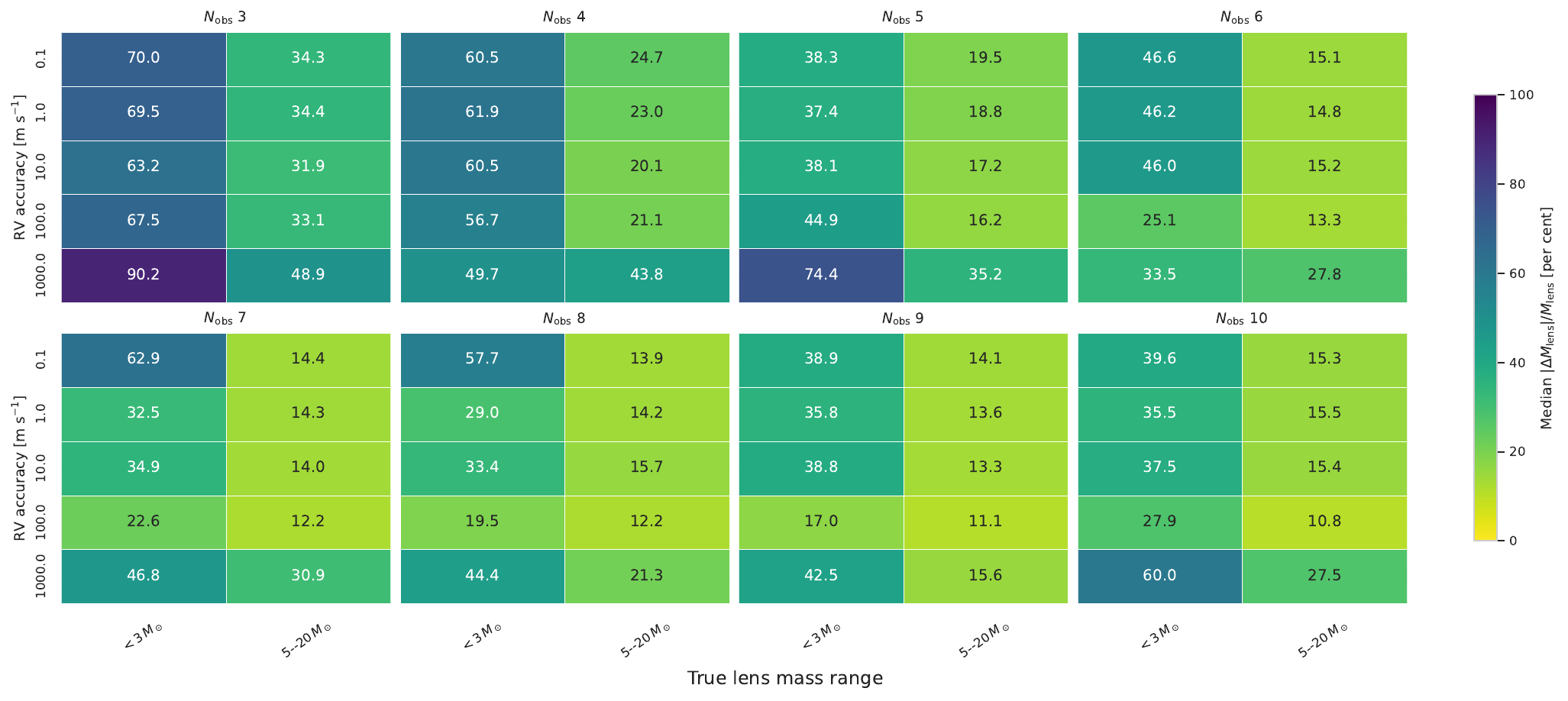}
    \caption{Median relative error for fitted lens mass scaled to the detectable population from population synthesis. Cell values give the median emulated $|\Delta M_\mathrm{lens}|/M_\mathrm{lens}$ in per cent for each observing strategy and instrumental RV accuracy. Columns within each panel give bins in true lens mass; the results are aggregated over orbital period. The population-synthesis sample contains no objects in the omitted lens-mass ranges.}
    \label{fig:population-weighted-rv-followup-heatmaps-relative-error-macc}
\end{figure*}

Figure~\ref{fig:population-weighted-rv-followup-heatmaps-relative-error-macc} shows the median relative errors for lens mass. Larger lens masses are recovered with smaller errors than smaller masses, and errors are significantly worse for $N_\mathrm{obs} < 5$, especially for lower lens masses.

These results follow the same trends as the unscaled results, showing that the unscaled results are informative about the potential for recovering the properties of the real population, even though the underlying parameter distribution is not physically representative.

\clearpage
\clearpage
\section{Conclusions} \label{sec:conclusions}
Previous works have shown that self-lensing systems can be confirmed from photometry alone, or with a combination of photometry and radial velocity measurements, by the discovery of five relatively short period white dwarf lenses. We show it is possible to characterise lenses across the full compact object mass range, including for very long orbital periods where only one self-lensing event is detectable.

We have presented a proof-of-concept for a method of scheduling a small number of RV follow-up observations of self-lensing candidates with previously-observed photometric lensing signals. We show that with only a small number of RV measurements it is possible to determine the broad compact object class in many cases and to measure the lens mass to an acceptable relative error. This can then inform prioritisation of further observations, if needed. With this in mind, our recommendation for following up on photometrically-detected self-lensing candidates is a staged follow-up strategy, where all candidates are observed a small number of times to determine which ones are truly binaries and what the lens class is in those cases. A second stage of follow-up observations can then concentrate on the true binaries to ascertain the lens mass to a higher precision.

We show that the highest mass lenses tend to have the lowest relative errors, even with large orbital periods. We also show that the number of observations matters more than RV precision in many cases. This is helpful when it comes to planning follow-up observations, as it widens the pool of facilities which can provide useful RV measurements. It is also promising that we are able to constrain the simulation parameters even for very long period systems. This means it is worthwhile following up on single-event detections as well as those which are seen to be periodic.

We show that a targeted radial velocity observing campaign, carefully timing observations to distinguish between possible solutions, is effective in recovering object properties with a minimal number of observations. It is possible to compute such a targeted schedule by fitting photometry alone.

Combining parameter recovery rates with previous population synthesis results suggests that it is possible to confirm the existence of a substantial number of self-lensing systems in surveys such as TESS, ZTF, LSST, and PLATO. The estimates presented here are optimistic upper limits for confirmed detections under ideal conditions, but bode well for our ability to find self-lensing systems in these and other surveys.

\section*{Acknowledgements}
The authors acknowledge the use of the Codex coding agent to assist in creating and debugging the code underlying these simulations.

This work was supported by the Science \& Technology Facilities Council [Grant Reference No ST/Y001699/1].

\section*{Data Availability}

The Python code and \LaTeX~source for this paper are available on GitHub (\href{https://github.com/adammcmaster/self_lensing_radial_velocity}{\texttt{adammcmaster/self\_lensing\_radial\_velocity}}), including scripts for generating all simulated data.



\bibliographystyle{mnras}
\bibliography{references}

@article{wright_radial_2005,
	title = {Radial {Velocity} {Jitter} in {Stars} from the {California} and {Carnegie} {Planet} {Search} at {Keck} {Observatory}},
	volume = {117},
	issn = {0004-6280, 1538-3873},
	url = {http://iopscience.iop.org/article/10.1086/430369},
	doi = {10.1086/430369},
	language = {en},
	number = {833},
	urldate = {2026-10-09},
	journal = {Publications of the Astronomical Society of the Pacific},
	author = {Wright, J. T.},
	month = jul,
	year = {2005},
	pages = {657--664},
}

@article{pepe_espresso_2021,
	title = {{ESPRESSO} at {VLT}: {On}-sky performance and first results},
	volume = {645},
	copyright = {https://www.edpsciences.org/en/authors/copyright-and-licensing},
	issn = {0004-6361, 1432-0746},
	shorttitle = {{ESPRESSO} at {VLT}},
	url = {https://www.aanda.org/10.1051/0004-6361/202038306},
	doi = {10.1051/0004-6361/202038306},
	urldate = {2026-10-09},
	journal = {Astronomy \& Astrophysics},
	author = {Pepe, F. and Cristiani, S. and Rebolo, R. and Santos, N. C. and Dekker, H. and Cabral, A. and Di Marcantonio, P. and Figueira, P. and Lo Curto, G. and Lovis, C. and Mayor, M. and Mégevand, D. and Molaro, P. and Riva, M. and Zapatero Osorio, M. R. and Amate, M. and Manescau, A. and Pasquini, L. and Zerbi, F. M. and Adibekyan, V. and Abreu, M. and Affolter, M. and Alibert, Y. and Aliverti, M. and Allart, R. and Allende Prieto, C. and Álvarez, D. and Alves, D. and Avila, G. and Baldini, V. and Bandy, T. and Barros, S. C. C. and Benz, W. and Bianco, A. and Borsa, F. and Bourrier, V. and Bouchy, F. and Broeg, C. and Calderone, G. and Cirami, R. and Coelho, J. and Conconi, P. and Coretti, I. and Cumani, C. and Cupani, G. and D’Odorico, V. and Damasso, M. and Deiries, S. and Delabre, B. and Demangeon, O. D. S. and Dumusque, X. and Ehrenreich, D. and Faria, J. P. and Fragoso, A. and Genolet, L. and Genoni, M. and Génova Santos, R. and González Hernández, J. I. and Hughes, I. and Iwert, O. and Kerber, F. and Knudstrup, J. and Landoni, M. and Lavie, B. and Lillo-Box, J. and Lizon, J.-L. and Maire, C. and Martins, C. J. A. P. and Mehner, A. and Micela, G. and Modigliani, A. and Monteiro, M. A. and Monteiro, M. J. P. F. G. and Moschetti, M. and Murphy, M. T. and Nunes, N. and Oggioni, L. and Oliveira, A. and Oshagh, M. and Pallé, E. and Pariani, G. and Poretti, E. and Rasilla, J. L. and Rebordão, J. and Redaelli, E. M. and Santana Tschudi, S. and Santin, P. and Santos, P. and Ségransan, D. and Schmidt, T. M. and Segovia, A. and Sosnowska, D. and Sozzetti, A. and Sousa, S. G. and Spanò, P. and Suárez Mascareño, A. and Tabernero, H. and Tenegi, F. and Udry, S. and Zanutta, A.},
	month = jan,
	year = {2021},
	pages = {A96},
}

@article{gaulme_kic_2022,
	title = {{KIC} 7955301: {A} hierarchical triple system with eclipse timing variations and an oscillating red giant},
	volume = {668},
	copyright = {https://creativecommons.org/licenses/by/4.0},
	issn = {0004-6361, 1432-0746},
	shorttitle = {{KIC} 7955301},
	url = {https://www.aanda.org/10.1051/0004-6361/202244373},
	doi = {10.1051/0004-6361/202244373},
	urldate = {2026-10-09},
	journal = {Astronomy \& Astrophysics},
	author = {Gaulme, Patrick and Borkovits, Tamás and Appourchaux, Thierry and Pavlovski, Krešimir and Spada, Federico and Gehan, Charlotte and Ong, Joel and Miglio, Andrea and Tkachenko, Andrew and Mosser, Benoît and Vrard, Mathieu and Benbakoura, Mansour and Drew Chojnowski, Stephen and Perkins, Jean and Hedlund, Anne and Jackiewicz, Jason},
	month = dec,
	year = {2022},
	pages = {A173},
}

@BOOK{eddington1926,
       author = {{Eddington}, A.~S.},
        title = "{The Internal Constitution of the Stars}",
    publisher = "{Cambridge University Press}",
         year = 1926,
       adsurl = {https://ui.adsabs.harvard.edu/abs/1926ics..book.....E},
    
}

@article{kallinger_connection_2014,
  title     = {The connection between stellar granulation and oscillation as seen by the {Kepler} mission},
  volume    = {570},
  issn      = {0004-6361},
  url       = {https://ui.adsabs.harvard.edu/abs/2014A&A...570A..41K},
  doi       = {10.1051/0004-6361/201424313},
  urldate   = {2026-08-17},
  journal   = {Astronomy and Astrophysics},
  publisher = {EDP},
  author    = {Kallinger, T. and De Ridder, J. and Hekker, S. and Mathur, S. and Mosser, B. and Gruberbauer, M. and García, R. A. and Karoff, C. and Ballot, J.},
  month     = oct,
  year      = {2014},
  note      = {ADS Bibcode: 2014A\&A...570A..41K},
  pages     = {A41}
}

@article{sorabella_modeling_2022,
  title    = {Modeling {Long}-term {Variability} in {Stellar}-compact {Object} {Binary} {Systems} for {Mass} {Determinations}},
  volume   = {936},
  issn     = {0004-637X},
  url      = {https://dx.doi.org/10.3847/1538-4357/ac82b7},
  doi      = {10.3847/1538-4357/ac82b7},
  language = {en},
  number   = {1},
  urldate  = {2023-09-05},
  journal  = {The Astrophysical Journal},
  author   = {Sorabella, Nicholas M. and Bhattacharya, Sayantan and Laycock, Silas G. T. and Christodoulou, Dimitris M. and Massarotti, Alessandro},
  month    = aug,
  year     = {2022},
  pages    = {63}
}

@article{witt_can_1994,
  title    = {Can {Lensed} {Stars} {Be} {Regarded} as {Pointlike} for {Microlensing} by {MACHOs}?},
  volume   = {430},
  issn     = {0004-637X},
  url      = {https://ui.adsabs.harvard.edu/abs/1994ApJ...430..505W},
  doi      = {10.1086/174426},
  urldate  = {2023-02-27},
  journal  = {The Astrophysical Journal},
  author   = {Witt, Hans J. and Mao, Shude},
  month    = aug,
  year     = {1994},
  note     = {ADS Bibcode: 1994ApJ...430..505W},
  pages    = {505}
}

@article{wiktorowicz_predicting_2021,
  title   = {Predicting the {Self}-lensing {Population} in {Optical} {Surveys}},
  volume  = {507},
  issn    = {0035-8711},
  url     = {https://doi.org/10.1093/mnras/stab2135},
  doi     = {10.1093/mnras/stab2135},
  number  = {1},
  journal = {Monthly Notices of the Royal Astronomical Society},
  author  = {Wiktorowicz, Grzegorz and Middleton, Matthew and Khan, Norman and Ingram, Adam and Gandhi, Poshak and Dickinson, Hugh},
  month   = oct,
  year    = {2021},
  pages   = {374--384}
}

@article{lucy_gravity-darkening_1967,
  title   = {Gravity-{Darkening} for {Stars} with {Convective} {Envelopes}},
  volume  = {65},
  issn    = {0372-8331},
  url     = {https://ui.adsabs.harvard.edu/abs/1967ZA.....65...89L},
  urldate = {2026-07-27},
  journal = {Zeitschrift fur Astrophysik},
  author  = {Lucy, L. B.},
  month   = jan,
  year    = {1967},
  note    = {ADS Bibcode: 1967ZA.....65...89L},
  pages   = {89}
}

@article{masuda_self-lensing_2019,
  title     = {Self-lensing {Discovery} of a 0.2 {M} {White} {Dwarf} in an {Unusually} {Wide} {Orbit} around a {Sun}-like {Star}∗},
  volume    = {881},
  issn      = {2041-8205},
  url       = {https://dx.doi.org/10.3847/2041-8213/ab321b},
  doi       = {10.3847/2041-8213/ab321b},
  language  = {en},
  number    = {1},
  urldate   = {2024-08-11},
  journal   = {The Astrophysical Journal Letters},
  publisher = {The American Astronomical Society},
  author    = {Masuda, Kento and Kawahara, Hajime and Latham, David W. and Bieryla, Allyson and Kunitomo, Masanobu and MacLeod, Morgan and Aoki, Wako},
  month     = aug,
  year      = {2019},
  pages     = {L3}
}

@article{kruse_koi-3278_2014,
  title      = {{KOI}-3278: {A} {Self}-{Lensing} {Binary} {Star} {System}},
  volume     = {344},
  shorttitle = {{KOI}-3278},
  url        = {https://www.science.org/doi/10.1126/science.1251999},
  doi        = {10.1126/science.1251999},
  number     = {6181},
  urldate    = {2024-08-21},
  journal    = {Science},
  publisher  = {American Association for the Advancement of Science},
  author     = {Kruse, Ethan and Agol, Eric},
  month      = apr,
  year       = {2014},
  pages      = {275--277}
}

@article{kawahara_discovery_2018,
  title     = {Discovery of {Three} {Self}-lensing {Binaries} from {Kepler}},
  volume    = {155},
  issn      = {1538-3881},
  url       = {https://dx.doi.org/10.3847/1538-3881/aaaaaf},
  doi       = {10.3847/1538-3881/aaaaaf},
  language  = {en},
  number    = {3},
  urldate   = {2024-08-21},
  journal   = {The Astronomical Journal},
  publisher = {The American Astronomical Society},
  author    = {Kawahara, Hajime and Masuda, Kento and MacLeod, Morgan and Latham, David W. and Bieryla, Allyson and Benomar, Othman},
  month     = mar,
  year      = {2018},
  pages     = {144}
}

@article{yahalomi_mass_2019,
  title      = {The {Mass} of the {White} {Dwarf} {Companion} in the {Self}-lensing {Binary} {KOI}-3278: {Einstein} versus {Newton}},
  volume     = {880},
  issn       = {0004-637X, 1538-4357},
  shorttitle = {The {Mass} of the {White} {Dwarf} {Companion} in the {Self}-lensing {Binary} {KOI}-3278},
  url        = {https://iopscience.iop.org/article/10.3847/1538-4357/ab2649},
  doi        = {10.3847/1538-4357/ab2649},
  number     = {1},
  urldate    = {2026-08-03},
  journal    = {The Astrophysical Journal},
  author     = {Yahalomi, Daniel A. and Shvartzvald, Yossi and Agol, Eric and Shporer, Avi and Latham, David W. and Kruse, Ethan and Brewer, John M. and Buchhave, Lars A. and Fulton, Benjamin J. and Howard, Andrew W. and Isaacson, Howard and Petigura, Erik A. and Quinn, Samuel N.},
  month      = jul,
  year       = {2019},
  pages      = {33}
}

@article{loeb_periodic_2003,
  title    = {Periodic {Flux} {Variability} of {Stars} due to the {Reflex} {Doppler} {Effect} {Induced} by {Planetary} {Companions}},
  volume   = {588},
  issn     = {0004-637X, 1538-4357},
  url      = {https://iopscience.iop.org/article/10.1086/375551},
  doi      = {10.1086/375551},
  language = {en},
  number   = {2},
  urldate  = {2026-08-03},
  journal  = {The Astrophysical Journal},
  author   = {Loeb, Abraham and Gaudi, B. Scott},
  month    = may,
  year     = {2003},
  pages    = {L117--L120}
}

@article{zucker_beaming_2007,
  title      = {Beaming {Binaries}: {A} {New} {Observational} {Category} of {Photometric} {Binary} {Stars}},
  volume     = {670},
  issn       = {0004-637X, 1538-4357},
  shorttitle = {Beaming {Binaries}},
  url        = {https://iopscience.iop.org/article/10.1086/521389},
  doi        = {10.1086/521389},
  language   = {en},
  number     = {2},
  urldate    = {2026-08-03},
  journal    = {The Astrophysical Journal},
  author     = {Zucker, Shay and Mazeh, Tsevi and Alexander, Tal},
  month      = dec,
  year       = {2007},
  pages      = {1326--1330}
}

@article{van_kerkwijk_observations_2010,
  title   = {{OBSERVATIONS} {OF} {DOPPLER} {BOOSTING} {IN} {KEPLER} {LIGHT} {CURVES}},
  volume  = {715},
  issn    = {0004-637X, 1538-4357},
  url     = {https://iopscience.iop.org/article/10.1088/0004-637X/715/1/51},
  doi     = {10.1088/0004-637X/715/1/51},
  number  = {1},
  urldate = {2026-08-03},
  journal = {The Astrophysical Journal},
  author  = {Van Kerkwijk, Marten H. and Rappaport, Saul A. and Breton, René P. and Justham, Stephen and Podsiadlowski, Philipp and Han, Zhanwen},
  month   = may,
  year    = {2010},
  pages   = {51--58}
}

@article{faigler_photometric_2011,
  title      = {Photometric detection of non-transiting short-period low-mass companions through the beaming, ellipsoidal and reflection effects in {Kepler} and {CoRoT} light curves: {New} method of detecting binaries},
  volume     = {415},
  issn       = {00358711},
  shorttitle = {Photometric detection of non-transiting short-period low-mass companions through the beaming, ellipsoidal and reflection effects in {Kepler} and {CoRoT} light curves},
  url        = {https://academic.oup.com/mnras/article-lookup/doi/10.1111/j.1365-2966.2011.19011.x},
  doi        = {10.1111/j.1365-2966.2011.19011.x},
  language   = {en},
  number     = {4},
  urldate    = {2026-08-03},
  journal    = {Monthly Notices of the Royal Astronomical Society},
  author     = {Faigler, S. and Mazeh, T.},
  month      = aug,
  year       = {2011},
  pages      = {3921--3928}
}

@article{shao_population_2019,
  title    = {Population {Synthesis} of {Black} {Hole} {Binaries} with {Normal}-star {Companions}. {I}. {Detached} {Systems}},
  volume   = {885},
  issn     = {0004-637X, 1538-4357},
  url      = {https://iopscience.iop.org/article/10.3847/1538-4357/ab4816},
  doi      = {10.3847/1538-4357/ab4816},
  number   = {2},
  urldate  = {2026-08-06},
  journal  = {The Astrophysical Journal},
  author   = {Shao, Yong and Li, Xiang-Dong},
  month    = nov,
  year     = {2019},
  pages    = {151}
}

@article{wiktorowicz_noninteracting_2020,
  title     = {Noninteracting {Black} {Hole} {Binaries} with {Gaia} and {LAMOST}},
  volume    = {905},
  issn      = {0004-637X},
  url       = {https://ui.adsabs.harvard.edu/abs/2020ApJ...905..134W},
  doi       = {10.3847/1538-4357/abc699},
  urldate   = {2024-08-03},
  journal   = {The Astrophysical Journal},
  publisher = {IOP},
  author    = {Wiktorowicz, Grzegorz and Lu, Youjun and Wyrzykowski, {\L}ukasz and Zhang, Haotong and Liu, Jifeng and Justham, Stephen and Belczynski, Krzysztof},
  month     = dec,
  year      = {2020},
  note      = {ADS Bibcode: 2020ApJ...905..134W},
  pages     = {134}
}

@article{weller_mapping_2023,
  title     = {Mapping progenitors of binary black holes and neutron stars with binary population synthesis},
  volume    = {520},
  copyright = {https://academic.oup.com/journals/pages/open\_access/funder\_policies/chorus/standard\_publication\_model},
  issn      = {0035-8711, 1365-2966},
  url       = {https://academic.oup.com/mnras/article/520/1/935/6991439},
  doi       = {10.1093/mnras/stad181},
  language  = {en},
  number    = {1},
  urldate   = {2026-08-06},
  journal   = {Monthly Notices of the Royal Astronomical Society},
  author    = {Weller, Miqaela K and Johnson, Jennifer A},
  month     = jan,
  year      = {2023},
  pages     = {935--947}
}

@article{paczynski_gravitational_1986,
  title     = {Gravitational {Microlensing} by the {Galactic} {Halo}},
  volume    = {304},
  issn      = {0004-637X},
  url       = {https://ui.adsabs.harvard.edu/abs/1986ApJ...304....1P},
  doi       = {10.1086/164140},
  urldate   = {2024-08-14},
  journal   = {The Astrophysical Journal},
  publisher = {IOP},
  author    = {Paczynski, B.},
  month     = may,
  year      = {1986},
  note      = {ADS Bibcode: 1986ApJ...304....1P},
  pages     = {1}
}

@article{paczynski_gravitational_1996,
  title    = {{GRAVITATIONAL} {MICROLENSING} {IN} {THE} {LOCAL} {GROUP}},
  volume   = {34},
  issn     = {0066-4146, 1545-4282},
  url      = {https://www.annualreviews.org/doi/10.1146/annurev.astro.34.1.419},
  doi      = {10.1146/annurev.astro.34.1.419},
  language = {en},
  number   = {1},
  urldate  = {2026-08-06},
  journal  = {Annual Review of Astronomy and Astrophysics},
  author   = {Paczyński, Bohdan},
  month    = sep,
  year     = {1996},
  pages    = {419--459}
}

@article{mao_astrophysical_2012,
  title   = {Astrophysical applications of gravitational microlensing},
  volume  = {12},
  issn    = {1674-4527},
  url     = {https://iopscience.iop.org/article/10.1088/1674-4527/12/8/005},
  doi     = {10.1088/1674-4527/12/8/005},
  number  = {8},
  urldate = {2026-08-06},
  journal = {Research in Astronomy and Astrophysics},
  author  = {Mao, Shude},
  month   = aug,
  year    = {2012},
  pages   = {947--972}
}

@article{sahu_isolated_2022,
  title     = {An {Isolated} {Stellar}-mass {Black} {Hole} {Detected} through {Astrometric} {Microlensing}},
  volume    = {933},
  issn      = {0004-637X},
  url       = {https://dx.doi.org/10.3847/1538-4357/ac739e},
  doi       = {10.3847/1538-4357/ac739e},
  language  = {en},
  number    = {1},
  urldate   = {2023-07-27},
  journal   = {The Astrophysical Journal},
  publisher = {The American Astronomical Society},
  author    = {Sahu, Kailash C. and Anderson, Jay and Casertano, Stefano and Bond, Howard E. and Udalski, Andrzej and Dominik, Martin and Calamida, Annalisa and Bellini, Andrea and Brown, Thomas M. and Rejkuba, Marina and Bajaj, Varun and Kains, Noé and Ferguson, Henry C. and Fryer, Chris L. and Yock, Philip and Mróz, Przemek and Kozłowski, Szymon and Pietrukowicz, Paweł and Poleski, Radek and Skowron, Jan and Soszyński, Igor and Szymański, Michał K. and Ulaczyk, Krzysztof and Wyrzykowski, {\L}ukasz and Collaboration), (OGLE and Barry, Richard K. and Bennett, David P. and Bond, Ian A. and Hirao, Yuki and Silva, Stela Ishitani and Kondo, Iona and Koshimoto, Naoki and Ranc, Clément and Rattenbury, Nicholas J. and Sumi, Takahiro and Suzuki, Daisuke and Tristram, Paul J. and Vandorou, Aikaterini and Collaboration), (MOA and Beaulieu, Jean-Philippe and Marquette, Jean-Baptiste and Cole, Andrew and Fouqué, Pascal and Hill, Kym and Dieters, Stefan and Coutures, Christian and Dominis-Prester, Dijana and Bennett, Clara and Bachelet, Etienne and Menzies, John and Albrow, Michael and Pollard, Karen and Collaboration), (PLANET and Gould, Andrew and Yee, Jennifer C. and Allen, William and Almeida, Leonardo A. and Christie, Grant and Drummond, John and Gal-Yam, Avishay and Gorbikov, Evgeny and Jablonski, Francisco and Lee, Chung-Uk and Maoz, Dan and Manulis, Ilan and McCormick, Jennie and Natusch, Tim and Pogge, Richard W. and Shvartzvald, Yossi and Collaboration), ( μFUN and Jørgensen, Uffe G. and Alsubai, Khalid A. and Andersen, Michael I. and Bozza, Valerio and Novati, Sebastiano Calchi and Burgdorf, Martin and Hinse, Tobias C. and Hundertmark, Markus and Husser, Tim-Oliver and Kerins, Eamonn and Longa-Peña, Penelope and Mancini, Luigi and Penny, Matthew and Rahvar, Sohrab and Ricci, Davide and Sajadian, Sedighe and Skottfelt, Jesper and Snodgrass, Colin and Southworth, John and Tregloan-Reed, Jeremy and Wambsganss, Joachim and Wertz, Olivier and Consortium), (MiNDSTEp and Tsapras, Yiannis and Street, Rachel A. and Bramich, D. M. and Horne, Keith and Steele, Iain A. and Collaboration), (RoboNet},
  month     = jul,
  year      = {2022},
  pages     = {83}
}

@article{lam_isolated_2022,
  title     = {An {Isolated} {Mass}-gap {Black} {Hole} or {Neutron} {Star} {Detected} with {Astrometric} {Microlensing}},
  volume    = {933},
  issn      = {2041-8205},
  url       = {https://dx.doi.org/10.3847/2041-8213/ac7442},
  doi       = {10.3847/2041-8213/ac7442},
  language  = {en},
  number    = {1},
  urldate   = {2024-08-21},
  journal   = {The Astrophysical Journal Letters},
  publisher = {The American Astronomical Society},
  author    = {Lam, Casey Y. and Lu, Jessica R. and Udalski, Andrzej and Bond, Ian and Bennett, David P. and Skowron, Jan and Mróz, Przemek and Poleski, Radek and Sumi, Takahiro and Szymański, Michał K. and Kozłowski, Szymon and Pietrukowicz, Paweł and Soszyński, Igor and Ulaczyk, Krzysztof and Wyrzykowski, {\L}ukasz and Miyazaki, Shota and Suzuki, Daisuke and Koshimoto, Naoki and Rattenbury, Nicholas J. and Hosek, Matthew W. and Abe, Fumio and Barry, Richard and Bhattacharya, Aparna and Fukui, Akihiko and Fujii, Hirosane and Hirao, Yuki and Itow, Yoshitaka and Kirikawa, Rintaro and Kondo, Iona and Matsubara, Yutaka and Matsumoto, Sho and Muraki, Yasushi and Olmschenk, Greg and Ranc, Clément and Okamura, Arisa and Satoh, Yuki and Silva, Stela Ishitani and Toda, Taiga and Tristram, Paul J. and Vandorou, Aikaterini and Yama, Hibiki and Abrams, Natasha S. and Agarwal, Shrihan and Rose, Sam and Terry, Sean K.},
  month     = jul,
  year      = {2022},
  pages     = {L23}
}

@article{gaudi_microlensing_2012,
  title    = {Microlensing {Surveys} for {Exoplanets}},
  volume   = {50},
  issn     = {0066-4146, 1545-4282},
  url      = {https://www.annualreviews.org/doi/10.1146/annurev-astro-081811-125518},
  doi      = {10.1146/annurev-astro-081811-125518},
  language = {en},
  number   = {1},
  urldate  = {2026-08-06},
  journal  = {Annual Review of Astronomy and Astrophysics},
  author   = {Gaudi, B. Scott},
  month    = sep,
  year     = {2012},
  pages    = {411--453}
}

@article{bond_ogle_2004,
  title      = {{OGLE} 2003-{BLG}-235/{MOA} 2003-{BLG}-53: {A} {Planetary} {Microlensing} {Event}},
  volume     = {606},
  issn       = {0004-637X},
  shorttitle = {{OGLE} 2003-{BLG}-235/{MOA} 2003-{BLG}-53},
  url        = {https://iopscience.iop.org/article/10.1086/420928/meta},
  doi        = {10.1086/420928},
  language   = {en},
  number     = {2},
  urldate    = {2024-08-13},
  journal    = {The Astrophysical Journal},
  publisher  = {IOP Publishing},
  author     = {Bond, I. A. and Udalski, A. and Jaroszyński, M. and Rattenbury, N. J. and Paczyński, B. and Soszyński, I. and Wyrzykowski, L. and Szymański, M. K. and Kubiak, M. and Szewczyk, O. and Żebruń, K. and Pietrzyński, G. and Abe, F. and Bennett, D. P. and Eguchi, S. and Furuta, Y. and Hearnshaw, J. B. and Kamiya, K. and Kilmartin, P. M. and Kurata, Y. and Masuda, K. and Matsubara, Y. and Muraki, Y. and Noda, S. and Okajima, K. and Sako, T. and Sekiguchi, T. and Sullivan, D. J. and Sumi, T. and Tristram, P. J. and Yanagisawa, T. and Yock, P. C. M. and Collaborations, The MOA {and} OGLE},
  month      = apr,
  year       = {2004},
  pages      = {L155}
}

@article{maeder_light_1973,
  title     = {Light {Curves} of the {Gravitational} {Lens}-like {Action} for {Binaries} with {Degenerate} {Stars}},
  volume    = {26},
  issn      = {0004-6361},
  url       = {https://ui.adsabs.harvard.edu/abs/1973A&A....26..215M},
  urldate   = {2026-08-06},
  journal   = {Astronomy and Astrophysics},
  publisher = {EDP},
  author    = {Maeder, A.},
  month     = jul,
  year      = {1973},
  note      = {ADS Bibcode: 1973A\&A....26..215M},
  pages     = {215}
}

@article{agol_occultation_2002,
  title    = {Occultation and {Microlensing}},
  volume   = {579},
  issn     = {0004-637X, 1538-4357},
  url      = {https://iopscience.iop.org/article/10.1086/342880},
  doi      = {10.1086/342880},
  language = {en},
  number   = {1},
  urldate  = {2026-08-06},
  journal  = {The Astrophysical Journal},
  author   = {Agol, Eric},
  month    = nov,
  year     = {2002},
  pages    = {430--436}
}

@book{rasmussen_gaussian_2005,
  title     = {Gaussian {Processes} for {Machine} {Learning}},
  copyright = {http://creativecommons.org/licenses/by-nc-nd/4.0/},
  isbn      = {978-0-262-25683-4},
  url       = {https://direct.mit.edu/books/book/2320/Gaussian-Processes-for-Machine-Learning},
  doi       = {10.7551/mitpress/3206.001.0001},
  language  = {en},
  urldate   = {2026-08-17},
  publisher = {The MIT Press},
  author    = {Rasmussen, Carl Edward and Williams, Christopher K. I.},
  month     = nov,
  year      = {2005}
}

@misc{kjeldsen_amplitudes_1995,
  title      = {Amplitudes of stellar oscillations: the implications for asteroseismology.},
  shorttitle = {Amplitudes of stellar oscillations},
  url        = {https://ui.adsabs.harvard.edu/abs/1995A&A...293...87K},
  doi        = {10.48550/arXiv.astro-ph/9403015},
  urldate    = {2026-08-06},
  publisher  = {arXiv},
  author     = {Kjeldsen, H. and Bedding, T. R.},
  month      = jan,
  year       = {1995},
  note       = {ISSN: 0004-6361
                Volume: 293
                ADS Bibcode: 1995A\&A...293...87K}
}

@article{chaplin_asteroseismology_2013,
  title    = {Asteroseismology of {Solar}-{Type} and {Red}-{Giant} {Stars}},
  volume   = {51},
  issn     = {0066-4146},
  url      = {https://ui.adsabs.harvard.edu/abs/2013ARA&A..51..353C},
  doi      = {10.1146/annurev-astro-082812-140938},
  urldate  = {2026-08-06},
  journal  = {Annual Review of Astronomy and Astrophysics},
  author   = {Chaplin, William J. and Miglio, Andrea},
  month    = aug,
  year     = {2013},
  note     = {ADS Bibcode: 2013ARA\&A..51..353C},
  pages    = {353--392}
}

@article{foreman-mackey_exoplanet_2021,
  title      = {exoplanet: {Gradient}-based probabilistic inference for exoplanet data \& other astronomical time series},
  volume     = {6},
  shorttitle = {exoplanet},
  url        = {https://ui.adsabs.harvard.edu/abs/2021JOSS....6.3285F},
  doi        = {10.21105/joss.03285},
  urldate    = {2026-08-06},
  journal    = {The Journal of Open Source Software},
  author     = {Foreman-Mackey, Daniel and Luger, Rodrigo and Agol, Eric and Barclay, Thomas and Bouma, Luke and Brandt, Timothy and Czekala, Ian and David, Trevor and Dong, Jiayin and Gilbert, Emily and Gordon, Tyler and Hedges, Christina and Hey, Daniel and Morris, Brett and Price-Whelan, Adrian and Savel, Arjun},
  month      = jun,
  year       = {2021},
  note       = {ADS Bibcode: 2021JOSS....6.3285F},
  pages      = {3285}
}

@article{pereira_gaussian_2019,
  title     = {Gaussian process modelling of granulation and oscillations in red giant stars},
  volume    = {489},
  issn      = {0035-8711},
  url       = {https://ui.adsabs.harvard.edu/abs/2019MNRAS.489.5764P},
  doi       = {10.1093/mnras/stz2405},
  urldate   = {2026-08-06},
  journal   = {Monthly Notices of the Royal Astronomical Society},
  publisher = {OUP},
  author    = {Pereira, Filipe and Campante, Tiago L. and Cunha, Margarida S. and Faria, João P. and Santos, Nuno C. and Barros, Susana C. C. and Demangeon, Olivier and Kuszlewicz, James S. and Corsaro, Enrico},
  month     = nov,
  year      = {2019},
  note      = {ADS Bibcode: 2019MNRAS.489.5764P},
  pages     = {5764--5774}
}

@article{brewer_gaussian_2009,
  title     = {Gaussian process modelling of asteroseismic data},
  volume    = {395},
  issn      = {0035-8711},
  url       = {https://ui.adsabs.harvard.edu/abs/2009MNRAS.395.2226B},
  doi       = {10.1111/j.1365-2966.2009.14679.x},
  urldate   = {2026-08-06},
  journal   = {Monthly Notices of the Royal Astronomical Society},
  publisher = {OUP},
  author    = {Brewer, B. J. and Stello, D.},
  month     = jun,
  year      = {2009},
  note      = {ADS Bibcode: 2009MNRAS.395.2226B},
  pages     = {2226--2233}
}

@article{nicholson_quasi-periodic_2022,
  title      = {Quasi-periodic {Gaussian} processes for stellar activity: {From} physical to kernel parameters},
  volume     = {515},
  issn       = {0035-8711},
  shorttitle = {Quasi-periodic {Gaussian} processes for stellar activity},
  url        = {https://ui.adsabs.harvard.edu/abs/2022MNRAS.515.5251N},
  doi        = {10.1093/mnras/stac2097},
  urldate    = {2026-08-06},
  journal    = {Monthly Notices of the Royal Astronomical Society},
  publisher  = {OUP},
  author     = {Nicholson, B. A. and Aigrain, S.},
  month      = oct,
  year       = {2022},
  note       = {ADS Bibcode: 2022MNRAS.515.5251N},
  pages      = {5251--5266}
}

@article{haywood_planets_2014,
  title      = {Planets and stellar activity: hide and seek in the {CoRoT}-7 system},
  volume     = {443},
  issn       = {0035-8711},
  shorttitle = {Planets and stellar activity},
  url        = {https://ui.adsabs.harvard.edu/abs/2014MNRAS.443.2517H},
  doi        = {10.1093/mnras/stu1320},
  urldate    = {2026-08-06},
  journal    = {Monthly Notices of the Royal Astronomical Society},
  publisher  = {OUP},
  author     = {Haywood, R. D. and Collier Cameron, A. and Queloz, D. and Barros, S. C. C. and Deleuil, M. and Fares, R. and Gillon, M. and Lanza, A. F. and Lovis, C. and Moutou, C. and Pepe, F. and Pollacco, D. and Santerne, A. and Ségransan, D. and Unruh, Y. C.},
  month      = sep,
  year       = {2014},
  note       = {ADS Bibcode: 2014MNRAS.443.2517H},
  pages      = {2517--2531}
}

@article{luger_starry_process_2021,
  title      = {starry\_process: {Interpretable} {Gaussian} processes for stellar light curves},
  volume     = {6},
  shorttitle = {starry\_process},
  url        = {https://ui.adsabs.harvard.edu/abs/2021JOSS....6.3071L},
  doi        = {10.21105/joss.03071},
  urldate    = {2026-08-06},
  journal    = {The Journal of Open Source Software},
  author     = {Luger, Rodrigo and Foreman-Mackey, Daniel and Hedges, Christina},
  month      = jul,
  year       = {2021},
  note       = {ADS Bibcode: 2021JOSS....6.3071L},
  pages      = {3071}
}

@article{aigrain_gaussian_2023,
  title    = {Gaussian {Process} {Regression} for {Astronomical} {Time} {Series}},
  volume   = {61},
  issn     = {0066-4146},
  url      = {https://ui.adsabs.harvard.edu/abs/2023ARA&A..61..329A},
  doi      = {10.1146/annurev-astro-052920-103508},
  urldate  = {2026-08-06},
  journal  = {Annual Review of Astronomy and Astrophysics},
  author   = {Aigrain, Suzanne and Foreman-Mackey, Daniel},
  month    = aug,
  year     = {2023},
  note     = {ADS Bibcode: 2023ARA\&A..61..329A},
  pages    = {329--371}
}

@article{koen_how_2024,
    title = {How many random observations are needed for good phase coverage of a periodic source?},
    volume = {57},
    issn = {0922-6435, 1572-9508},
    url = {https://link.springer.com/10.1007/s10686-024-09931-1},
    doi = {10.1007/s10686-024-09931-1},
    language = {en},
    number = {2},
    urldate = {2026-08-20},
    journal = {Experimental Astronomy},
    author = {Koen, Chris},
    month = apr,
    year = {2024},
    pages = {12},
}

@misc{foreman-mackey_dfmtinygp_2026,
    title = {dfm/tinygp: {The} tiniest of {Gaussian} {Process} libraries},
    copyright = {MIT License},
    shorttitle = {dfm/tinygp},
    url = {https://zenodo.org/doi/10.5281/zenodo.19035246},
    doi = {10.5281/ZENODO.19035246},
    urldate = {2026-09-03},
    publisher = {Zenodo},
    author = {Foreman-Mackey, Daniel and Weixiang Yu and Yadav, Sachin and Becker, McCoy Reynolds and Caplar, Neven and Huppenkothen, Daniela and Killestein, Thomas and Tronsgaard, René and Rashid, Theo and Schmerler, Steve},
    month = mar,
    year = {2026},
}

@misc{dixon_cobalt-blue_2026,
    title = {{COBALT}-{BLUE}: a fast and accurate model for gravitational self-lensing binary systems},
    shorttitle = {{COBALT}-{BLUE}},
    url = {http://arxiv.org/abs/2609.21708},
    doi = {10.48550/arXiv.2609.21708},
    urldate = {2026-09-21},
    publisher = {arXiv},
    author = {Dixon, Danny J. and Ingram, Adam and Middleton, Matthew J. and McMaster, Adam and Wiktorowicz, Grzegorz and Dickinson, Hugh and Crossland, Allison and Bellm, Eric C.},
    month = sep,
    year = {2026},
    note = {arXiv:2609.21708 [astro-ph.HE]},
}

@article{gonzalez_clustering_1985,
    title = {Clustering to minimize the maximum intercluster distance},
    volume = {38},
    copyright = {https://www.elsevier.com/tdm/userlicense/1.0/},
    issn = {03043975},
    url = {https://linkinghub.elsevier.com/retrieve/pii/0304397585902245},
    doi = {10.1016/0304-3975(85)90224-5},
    language = {en},
    urldate = {2026-10-06},
    journal = {Theoretical Computer Science},
    author = {Gonzalez, Teofilo F.},
    year = {1985},
    pages = {293--306},
}

@article{lloyd_least_1982,
    title = {Least squares quantization in {PCM}},
    volume = {28},
    copyright = {https://ieeexplore.ieee.org/Xplorehelp/downloads/license-information/IEEE.html},
    issn = {0018-9448, 1557-9654},
    url = {https://ieeexplore.ieee.org/document/1056489/},
    doi = {10.1109/TIT.1982.1056489},
    number = {2},
    urldate = {2026-10-06},
    journal = {IEEE Transactions on Information Theory},
    author = {Lloyd, S.},
    month = mar,
    year = {1982},
    pages = {129--137},
}




\appendix

\section{Photometric model} \label{sec:phot-model}
As with the full COBALT-BLUE model, we include three photometric effects: self-lensing, ellipsoidal variations, and Doppler boosting. We use the orbital phase convention described in Section~\ref{sec:phot-sims}, with the phase and true anomaly defined in Section~\ref{sec:rv-sims}.

The source-lens distance is calculated according to the phase angle at a given time, meaning the source-lens distance changes during the microlensing event according to the eccentricity. Strictly speaking, the Einstein ring radius should also change as a function of orbital phase, but this effect is expected to be small and so is not accounted for in the present work. Instead, the radius at $t_0$ is used for all calculations.

We use the Einstein radius and impact-parameter convention of \cite{wiktorowicz_predicting_2021}, generalised from the circular-orbit case by replacing the orbital separation with the instantaneous Keplerian lens-source separation,
\begin{equation}
    R_\mathrm{E} = \left(\frac{4 G M_\mathrm{lens} r(t_0) \sin i}{c^2}\right)^{1/2},
    \quad
    b = \frac{a |\cos i|}{R_\mathrm{E}},
    \quad
    b_\mathrm{max} = 1 + \frac{R_\star}{R_\mathrm{E}},
\end{equation}
where $r(t_0)$ is the lens-source separation at the lensing peak. For a circular orbit, $r(t_0)=a$ and this reduces to the expression used by \cite{wiktorowicz_predicting_2021}. The true anomaly, $\nu(t)$, is obtained from Kepler's equation using the mean anomaly $\mathcal{M}(t)$ and eccentric anomaly $E(t)$,
\begin{align}
    \mathcal{M}(t) & = \mathcal{M}_0 + 2\pi\frac{t - t_0}{P_\mathrm{orb}}, \\
    \mathcal{M}(t) & = E(t) - e\sin E(t),                                  \\
    \nu(t)         & = 2\arctan2\left[
        \sqrt{1+e}\sin\frac{E(t)}{2},
        \sqrt{1-e}\cos\frac{E(t)}{2}
        \right],                                                           \\
    P_\mathrm{orb} & =
    2\pi\left(\frac{a^3}{G(M_\mathrm{lens} + M_\star)}\right)^{1/2},
\end{align}
with $\mathcal{M}_0$ chosen such that $\phi(t_0)=0$. $\arctan2$ is the standard quadrant-aware inverse tangent. The true anomaly at the lensing peak is $\nu_0=-\omega$, so the corresponding eccentric anomaly and mean anomaly are
\begin{align}
    E_0           & = 2\arctan2\left[
        \sqrt{1-e}\sin\frac{\nu_0}{2},
        \sqrt{1+e}\cos\frac{\nu_0}{2}
        \right],                        \\
    \mathcal{M}_0 & = E_0 - e\sin E_0 .
\end{align}
For this orbital phase, the true orbital lens-source separation is
\begin{equation}
    r[\phi(t)] =
    \frac{a(1 - e^2)}
    {1 + e\cos[\phi(t) - \omega]}.
\end{equation}
The projected lens-source separation in Einstein-radius units is
\begin{equation}
    u(t) = \frac{1}{R_\mathrm{E}}
    \left[
        \left(r[\phi(t)] \sin\phi(t)\right)^2
        + \left(b R_\mathrm{E}\right)^2
        \right]^{1/2},
\end{equation}
where $u(t)$ is dimensionless. The self-lensing magnification is calculated with the integrated Witt--Mao finite-source formalism \citep{witt_can_1994}. The implementation evaluates the closed-form uniform-disc magnification,
\begin{equation}
    A_\mathrm{WM}(u, \rho_\star) =
    \frac{
        C_E E(m) + C_K K(m) + C_\Pi \Pi(n, m)
    }{\pi},
\end{equation}
where $K$, $E$, and $\Pi$ are complete elliptic integrals of the first, second, and third kind. The coefficients $C_E$, $C_K$, and $C_\Pi$, and the elliptic-integral arguments $m$ and $n$, are functions only of the projected lens-source separation and $\rho_\star = R_\star / R_\mathrm{E}$. The disc-integrated magnification used in the simulated light curves is then evaluated by quadrature over a linearly limb-darkened stellar disc,
\begin{align}
    A_\mathrm{SL}(t) & =
    \frac{
        \int_0^1
        A_\mathrm{WM}\left[u(t), \rho\rho_\star\right]
        I(\mu)\,2\rho\,\mathrm{d}\rho
    }{
        \int_0^1 I(\mu)\,2\rho\,\mathrm{d}\rho
    },
    \nonumber                             \\
    \mu              & = \sqrt{1-\rho^2},
    \quad
    I(\mu) = 1 - \gamma(1-\mu),
\end{align}


where $\gamma=0.6$ is the fixed linear limb-darkening coefficient. This is the value corresponding to Eddington's grey atmosphere \citep{eddington1926}. It is a fiducial approximation; in analyses of individual candidates, the coefficient should instead be interpolated from atmosphere models using the measured stellar parameters.

Calculation of ellipsoidal variations and Doppler boosting are adapted from \cite{sorabella_modeling_2022}, with the orbital separation and radial velocity generalised to the eccentric case. The ellipsoidal-variation term is
\begin{align}
    A_\mathrm{EV}(t) & =
    -\beta
    \frac{M_\mathrm{lens}}{M_\star}
    \left(\frac{R_\star}{r[\phi(t)]}\right)^3
    \sin^2 i \cos 2\phi(t),
    \nonumber                                                           \\
    \beta            & = 0.15\,\frac{(1 + g)(15 + \gamma)}{3 - \gamma},
\end{align}
where $\beta$ is the ellipsoidal-variation coefficient and $g=0.32$ is the gravity-darkening coefficient for a Sun-like star \citep{lucy_gravity-darkening_1967}.

The Doppler-boosting term is
\begin{equation}
    A_\mathrm{DB}(t) =
    (3 - \alpha)\frac{v_r(t)}{c},
    \quad
    \alpha =
    \frac{\exp(x)(3 - x) - 3}{\exp(x) - 1},
    \quad
    x = \frac{h\nu_\mathrm{TESS}}{k_\mathrm{B}T_\mathrm{eff}},
\end{equation}
where $\nu_\mathrm{TESS}$ is the TESS-band effective frequency and $v_r(t)$ is the source radial velocity. The combined model flux is
\begin{equation}
    F(t) = F_0 A_\mathrm{SL}(t)\left[1 + A_\mathrm{EV}(t) + A_\mathrm{DB}(t)\right],
\end{equation}
which makes the simplifying assumption that the ellipsoidal variations and Doppler boosting are of small amplitude.


\bsp	
\label{lastpage}
\end{document}